\documentclass[journal,10pt]{IEEEtran}

\usepackage{subfigure}
\usepackage{multirow,multicol}
\usepackage{amsmath}
\usepackage{amsthm}
\usepackage{epsfig}
\usepackage{url}
\usepackage{balance}
\usepackage{cite}
\usepackage{color}
\usepackage[numbers,sort&compress]{natbib}
\newtheorem{theorem}{\textbf{Theorem}}
\theoremstyle{plain}

\usepackage[cmintegrals]{newtxmath}

\begin{document}

\title{Analysis on Computation-Intensive Status Update in Mobile Edge Computing}
%
%

\author{Qiaobin~Kuang,
        Jie~Gong,~\IEEEmembership{Member,~IEEE},
        Xiang~Chen,~\IEEEmembership{Member,~IEEE},
        and~Xiao~Ma,~\IEEEmembership{Member,~IEEE}
\thanks{Part of this paper was presented at the 11th International Conference on Wireless Communications and Signal Processing, Xi'an, China, Oct. 2019 \cite{Kuang2019Age}.}
\thanks{Q. Kuang and X. Chen are with the School of Electronics and Information Technology, Sun Yat-sen University, Guangzhou 510006, China.}
\thanks{J. Gong and X. Ma are with the School of Data and Computer Science, Sun Yat-sen University, Guangzhou 510006, China. J. Gong is the corresponding author (email: gongj26@mail.sysu.edu.cn).}}

\markboth{IEEE Transactions on Vehicular Technology,~Vol.~XX, No.~XX, XXX~2020}
{}
\maketitle

\begin{abstract}
In status update scenarios, the freshness of information is measured in terms of age-of-information (AoI), which essentially reflects the timeliness for real-time applications to transmit status update messages to a remote controller.
For some applications, computational expensive and time consuming data processing is inevitable for status information of messages to be displayed.
Mobile edge servers are equipped with adequate computation resources and they are placed close to users. Thus, mobile edge computing (MEC) can be a promising technology to reduce AoI for computation-intensive messages.
In this paper, we study the AoI for computation-intensive messages with MEC, and consider three computing schemes: local computing, remote computing at the MEC server, and partial computing, i.e., some part of computing tasks are performed locally, and the rest is executed at the MEC server. Zero-wait policy is adopted in all three schemes. Specifically, in local computing, a new message is generated immediately after the previous one is revealed by computing.
 While in remote computing and partial computing, a new message is generated once the previous one is received by the remote MEC server. With infinite queue size and exponentially distributed transmission time,
 closed-form average AoI for exponentially distributed computing time is derived for the three computing schemes.
For deterministic computing time, the average AoI is analyzed numerically. Simulation results show that by carefully partitioning the computing tasks, the average AoI in partial computing is the smallest compared to local computing and remote computing. The results also indicate numerically the conditions on which remote computing attains smaller average AoI compared with local computing.
\end{abstract}


\IEEEpeerreviewmaketitle

\section{Introduction}
In recent years, various kinds of real-time applications such as ads bidding, stocks forecast, weather monitoring, and social networks have become a focus of attention.
These applications have high requirement in the freshness of status information for making accurate decision. The freshness of data can be measured by age-of-information (AoI) \cite{Kaul2011Minimizing,Kaul2012Real}.
It is defined as the time elapsed since the latest delivery of the update was generated.

AoI has attracted many researchers in academic. AoI was firstly proposed in \cite{Kaul2011Minimizing, Kaul2012Real} as a metric of the information freshness at the target node.
In \cite{Kaul2012Real}, the authors obtained a general result for extensive service systems with the update messages served with first-come-first-served (FCFS) principle, and specifically considered $M/M/1$, $M/D/1$ and $D/M/1$ standard queuing models.
In \cite{Roy2012Real}, status updating from multiple sources was analyzed. In \cite{Ahmed2017The} and \cite{Talak2017Minimizing}, minimizing AoI for multi-hop wireless networks with interference-free networks and general interference constrains were considered, respectively. The above references focus on the update messages stochastically generated at the source. Thus, the message has to wait in the queue when the server is busy. Thus AoI may increase due to the queuing delay. A just-in-time policy was proposed in \cite{Yates2015Lazy} to solve the problem according to the knowledge of the system state, for example, to generate messages only when the server is idle. The policy is also called zero-wait policy \cite{Sun2017Update} or the work-conserving policy \cite{Kleinrock1975Queueing}. The authors in \cite{Sun2017Update} also noticed that there can be better polices other than the zero-wait policy in many scenarios. In \cite{Costa2016On}, peak age was taken as a new measurement of the freshness of information because of its analytically convenience. Recently, some researchers are devoted to developing new tools for AoI analysis in networks. In particular, ref. \cite{Roy2018Age} explicitly calculated the average age over a multi-hop network of preemptive servers by using a stochastic hybrid system (SHS). And in \cite{Yates2018Age}, the authors applied SHS in the analysis of the temporal convergence of higher order AoI moments, and enable the moment generation function to characterize the stationary distribution of an AoI process in multi-hop networks.

The above papers only pay attention to  the influence of data transmission and queuing on AoI. However, the impact of data processing on AoI is non-negligible in some real-time applications.
Take autonomous driving as an example, when a status update is an image, it needs not only to be transmitted to the controller, but also to be processed to expose the embedded status information. Unfortunately, subject to the limited computational capacity of the local processor, data processing could be computational expensive and time consuming.
Mobile edge computing (MEC) can be a potential technique to solve the above problem for reducing the AoI of computation-intensive messages, since it has the ability to provide abundant cloud-like computing resource via integrated MEC servers deployed at the network edge such as access points and cellular base stations, as well as to cut down the response time in comparison with the centralized cloud \cite{Mao2017A,Mach2017Mobile}.  Motivated by this, we consider introducing MEC to process the computation-intensive message.

In MEC systems, the computing tasks can be offloaded to an MEC server. As the MEC server usually has sufficient computing capacity and is close to mobile users, offloading can greatly save the user's energy and reduce the computing time.
 For computation offloading, it is crucial to determine whether or not to offload and what and how much should be offloaded \cite{Zhang2012To}. The computation offloading decision is influenced by a number of parameters such as the QoS requirements to be met (e.g., in AoI minimization, energy efficiency maximization), as well as the capacities of processing nodes and the availability of radio resources for wireless packet transmission. And there are mainly two widely used computation task models, namely binary computation offloading as well as partial computation offloading \cite{Mao2017A,Mach2017Mobile}.
 For binary offloading, the task is inseparable because its tight integration or relatively simplicity, such as speech recognition and natural language translation. Thus, the whole computation task is performed either locally by the mobile user or offloaded to the MEC server.
For partial offloading, the computation task can be divided into  more than one part. Some parts are computed by mobile user locally and the rest are transmitted to and computed at the MEC server. Applications of partial offloading consist of multiple fine-grained processes/components, such as augmented reality and face detection.
If the composable components of the task are independent, the computing process can be executed both locally and remotely in parallel.
 In the literature, many works investigated computation offloading in MEC systems. Minimizing the execution delay is studied in \cite{Liu2016Delay,Mao2016Dynamic,Zhao2019Optimal,
 Ren2019Collaborative}.
 Under the constraint of execution delay, \cite{Sardellitti2015Joint,Wang2016Mobile,You2017Energy,Wang2018Joint} minimized the energy consumption and \cite{Rodrigues2018Cloudlets} maximized the system scalability.
  A balance between execution delay and energy consumption for computation offloading was considered in \cite{O2015Optimization, Chen2016Efficient,Dinh2017Offloading}.
 While in practical applications, completely parallel execution for the task-input bits may be unpractical, since the bit-wise correlation hinders arbitrarily division into different groups.
 In this paper, we consider the computation task generated at the information source to be computed in three ways: 1) local computing, where the task is computed as a whole at the local processor; 2) remote computing, where the task is computed as a whole at the MEC server; 3) partial computing, where the task is firstly computed at the local processor and then the output of the local processor is transferred to the MEC server for further computation. Note that in the third method, the size of the output obtained by means of local processing is less than the size of the message generated at the source node.

In terms of computation-intensive messages, apart from the two affects mentioned above, the message generation frequency and the delay caused by data transmission and queuing, data processing delay is also non-negligible for the research of AoI. For example, ref. \cite{Zhou2018Optimal} considered AoI minimization in two data processing scenes, the complicated initial feature extraction and classification in computer vision, as well as the optimization of sampling and updating processes in an Internet of things device's sampled physical process.
The authors in \cite{Zou2019Trading} considered a general analysis with packet management for average AoI and average peak AoI with a computation server and a transmission queue.
Ref. \cite{Alabbasi2018Joint} put forward new scheduling schemes for computing and network phases in vehicular networks by combining the computation and information freshness.
In \cite{Dong2019Timely}, the authors investigated bidirectional timely data exchanging between a fog node and a mobile user in a fog computing system. For a resource constrained edge cloud system, the authors in \cite{Zhong2019Age} considered a greedy traffic scheduling policy to minimize the overall age penalty of multiple users. In \cite{Arafa2019Timely}, a cloud computing status updating was studied with preemption policy. The authors in \cite{Song2019Age} proposed a new performance metric called age of task (AoT) to evaluate the temporal value of computation tasks and jointly considered task scheduling, computation offloading and energy consumption. Although the above papers considered data processing, they did not take advantage of MEC's short distance to the source node and sufficient computation resources.

In this paper, we concentrate on the average AoI for computation-intensive messages in an MEC system. We study the AoI performance with three computing strategies, including local computing, remote computing and partial computing. It has not been studied to the best of our knowledge. Zero-wait policy is applied to the three computing schemes.
Specifically, local computing generates the update message immediately after the computation of the last update message. For remote and partial computing, the generation of a new update message comes after the previous one's arrival at the MEC server. In the three computing schemes, the computing follows FCFS principle.
Assume the transmission time follows an exponential distribution, and consider exponentially distributed and deterministic computing time with infinite computing queue size. The main contributions of this article are summarized as follows:
\begin{itemize}
\item  We derive the closed-form average AoI for the three computing schemes with exponentially distributed computing time.
     We found that by carefully partitioning the computing task, partial computing performs the best compared with local computing and remote computing. And it is significantly better than remote computing when the ratio of transmission rate and remote computing rate is very small or very large.  If the transmission rate is small, the performance of local computing is the same as partial computing.

\item The average AoI with deterministic computing time is obtained numerically. Simulation results show that, with large local computing rate and with both small transmission rate and large one, local computing and partial computing have similar performance. While, with small local computing rate, remote computing outperforms local computing.

\item The influence of message size, required number of central processing unit (CPU) cycles, data rate, as well as computing capacity of the MEC server for data processing on the average AoI is studied by numerical simulations.
 It is found that remote computing does not always outperform local computing in terms of average AoI. We characterize numerically when remote computing should be adopted compared with local computing.
\end{itemize}

The rest of the paper is organized as follows. In the next section, the system model and the average AoI about the three computing schemes are presented. The analytic results for average AoI with exponentially distributed computing time and deterministic computing time are discussed in Section \uppercase\expandafter{\romannumeral3} and Section \uppercase\expandafter{\romannumeral4}, respectively.
And numerical analysis for exponentially distributed computing time are showed in Section \uppercase\expandafter{\romannumeral5}. This paper is concluded in Section \uppercase\expandafter{\romannumeral6}.

\section{System Model and Average AoI}
\begin{figure}
  \centering
  \subfigure[Local computing]{
    \label{fig:subfig:a} 
    \includegraphics[width=3.1in]{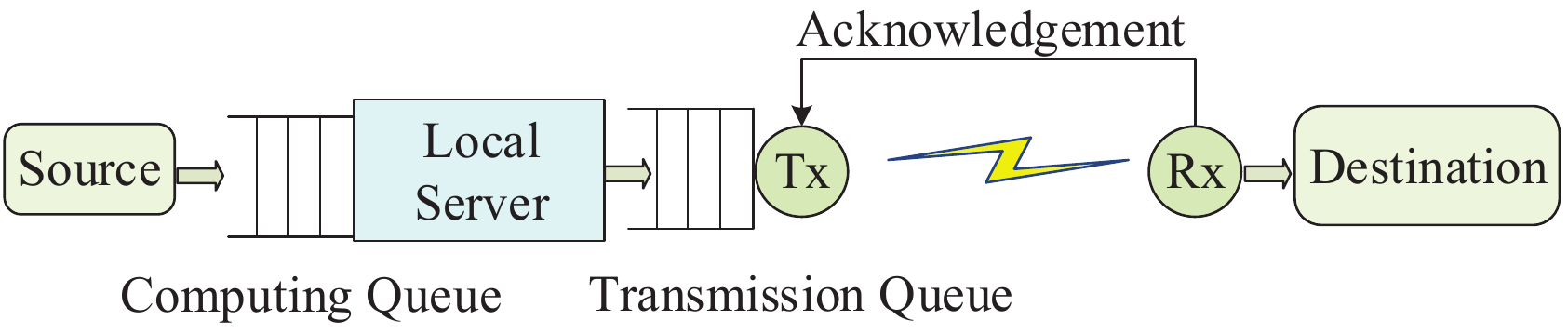}
  }
  \subfigure[Remote computing]{
    \label{fig:subfig:b} 
    \includegraphics[width=3.1in]{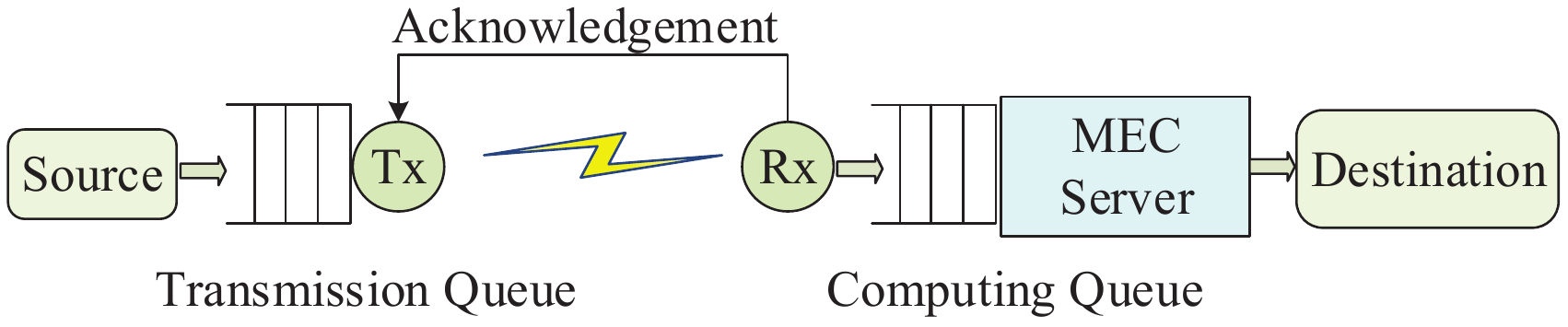}
  }
    \subfigure[Partial computing]{
    \label{fig:subfig:c} 
    \includegraphics[width=3.1in]{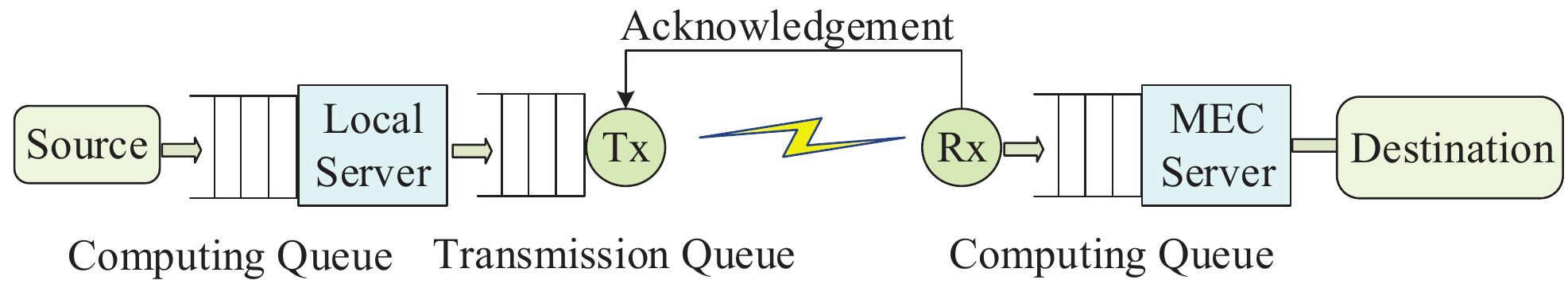}
  }
  \caption{System models}
  \label{fig:subfig} 
\end{figure}

Fig. \ref{fig:subfig} presents a status monitoring and control system for computation-intensive messages. Firstly, the source generates system status. Then, one of the local server and the remote server or both of them, will process system status. In the next, the target node receives the transmitted processed signal. During the procedure, it is vital to maintain the freshest processed status for the accuracy of control. Details of the three schemes of local computing,  remote computing and partial computing will be described in the following parts.
\subsection{Local computing, remote computing and partial computing model}
Since both the MEC server and the user have the computing ability, we compare three computing methods in this paper.
\subsubsection{Local Computing}
Depicted in Fig. \ref{fig:subfig:a}, this scheme allocates all computation-intensive data to local computation before sending the processed information to the target. In particular, the source firstly generates a status update message and then arrives at the computing queue. After the computation is completed by the local server, the computing result arrives at the transmission queue and then is transmitted to the destination node through the channel.
\subsubsection{Remote Computing}
In this scheme, the computation-intensive message is transmitted to the MEC server and then computed remotely, as illustrated in Fig. \ref{fig:subfig:b}. Particularly, the status update message generated by the source node is transmitted through the channel and arrives at the computing queue in the MEC server. Finally, the MEC server completes the computation of the status update message based on FCFS principle and sends the result to the destination node.
\subsubsection{Partial Computing}
As shown in Fig. \ref{fig:subfig:c}, the last scheme partially computes the computation-intensive data by the local server, and then sends intermediate computing result to the MEC server for further computing. Specifically, the local server partially processes the computation-intensive data, and the intermediate computing result enters the transmission queue. Then, the intermediate result is transmitted to the remote computing queue to wait for the MEC server to finish the rest computing part which is also based on the FCFS principle. When the computation is completely finished, the result can be sent to the destination node.
\subsection{Zero-Wait Message Generation Policy}
In this section, we present the three computing schemes with zero-wait message generation policy \cite{Sun2017Update}, a policy that new message is generated immediately after the last one completes its computing or transmission. Intuitively, the zero-wait policy attains good performance as the waiting in a queue is avoided. Other message generation policies will be considered in the future work.  The detailed zero-wait policies in the three schemes are given as follows.
\subsubsection{Local Computing}
\begin{figure}
  \centering
  \subfigure[Local computing]{
    \label{fig:subfig_evo:a} 
    \includegraphics[width=3.1in]{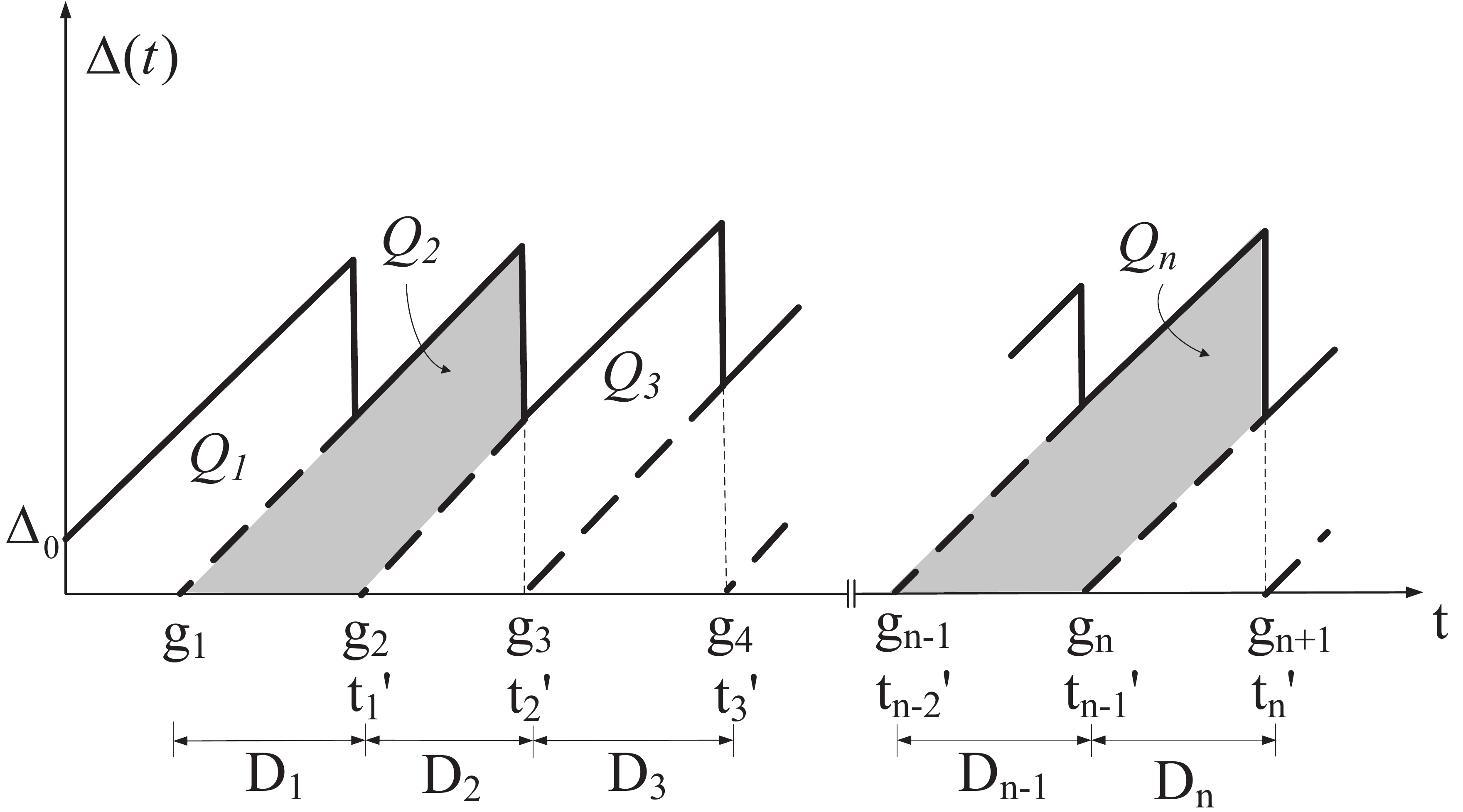}
  }
  \subfigure[Remote computing]{
    \label{fig:subfig_evo:b} 
    \includegraphics[width=3.1in]{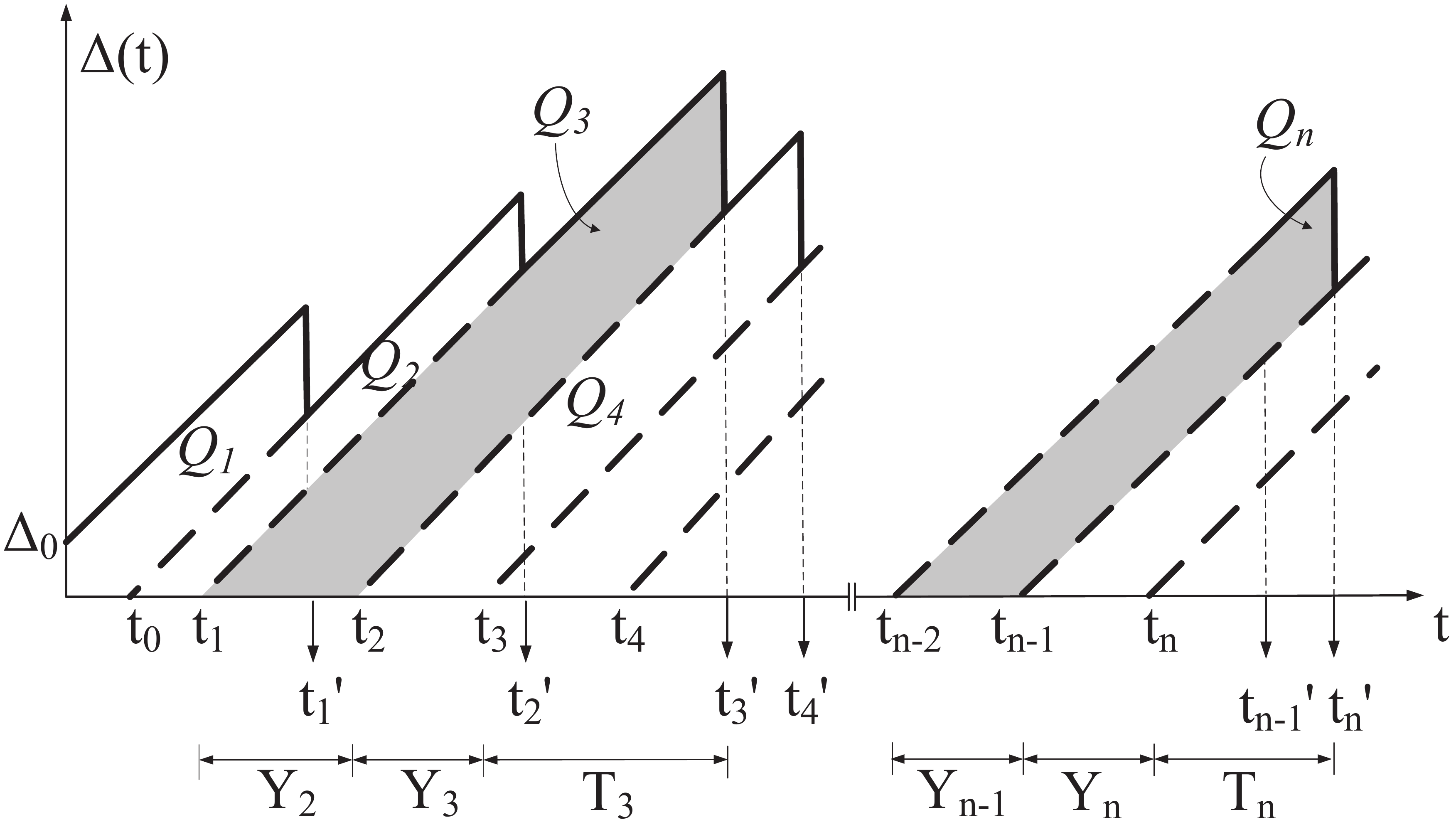}
  }
    \subfigure[Partial computing]{
    \label{fig:subfig_evo:c} 
    \includegraphics[width=3.1in]{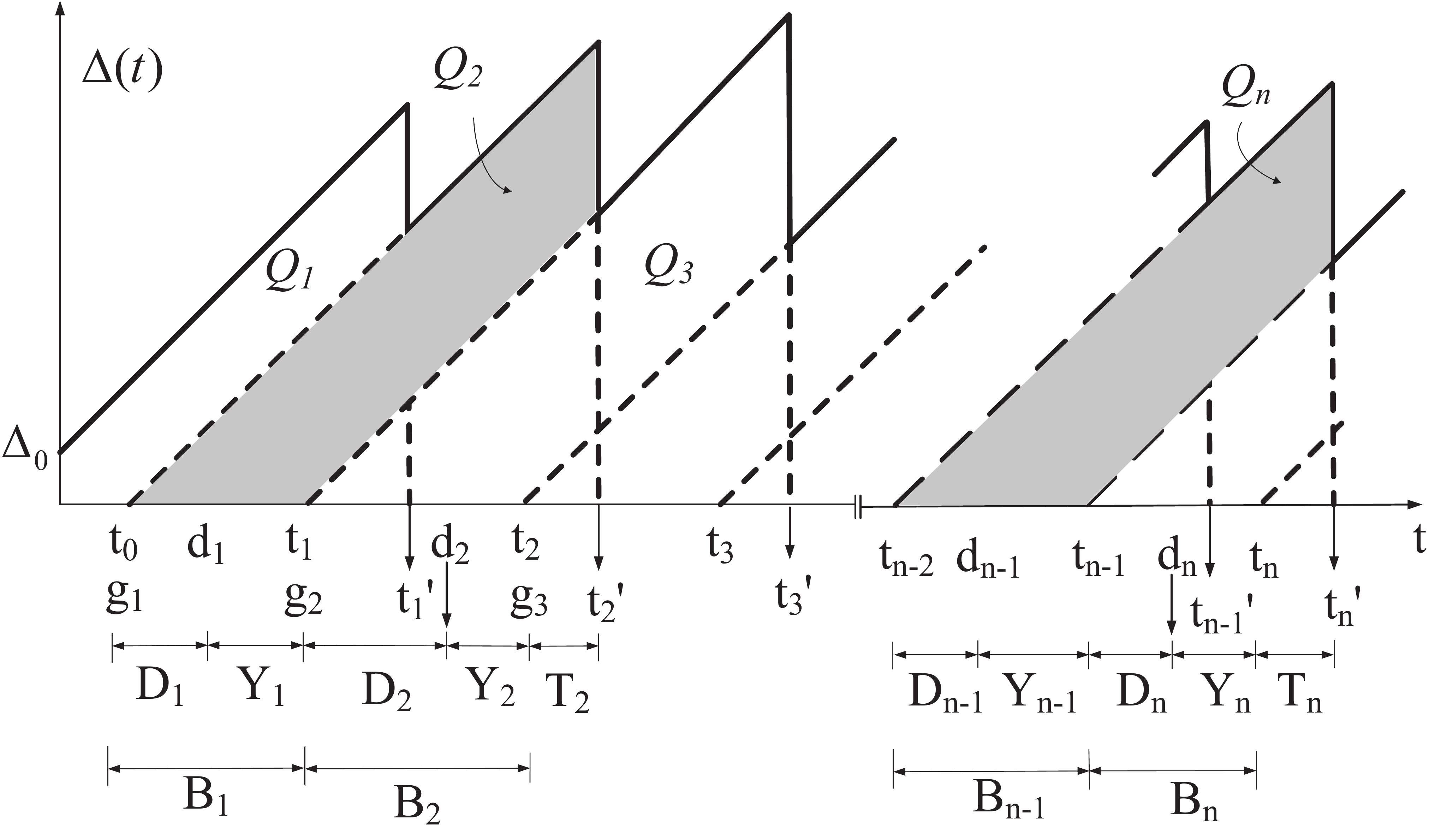}
  }
  \caption{Sawtooth curve - Examples of sample path for average AoI $\Delta(t)$ of the three computing schemes. }
  \label{fig:subfig_evo} 
\end{figure}

In local computing, a new status update message waits to be generated until the previous message is totally computed by the local server. Therefore, the computing queue is empty and the queuing delay is completely eliminated. Compared to the size of the original message, that of the computing result to be transmitted is negligible, thus the time to transmit the result to the target can be ignored when comparing with to the time for computing. Fig. \ref{fig:subfig_evo:a} illustrates the evolution of the AoI $\Delta(t)$ at the destination node for local computing under FCFS queuing, where $g_i$ denotes the generation time instant of the $i$-th status update message, $t_i^{'}$ is the time instant when message $i$-th is computed locally. When the computing is finished, the revealed status information is transmitted to the destination node. Therefore, the age drops suddenly at time $t_i^{'}$, and a new message is generated at that time.
\subsubsection{Remote Computing}
In remote computing, zero-wait policy means that once the receiver receives a status update message, it sends an acknowledgement signal to the source node, and a new status update message will immediately be generated in the source node and be transmitted. Due to the size of the acknowledgement signal is relatively much smaller compared to that of the status update message, the feedback time is ignored. With zero-wait policy, the queuing delay for transmission is zero. The delivered status update message waits in a queue before the MEC server, and will be served with FCFS principle. Fig. \ref{fig:subfig_evo:b} shows the change of the AoI $\Delta(t)$ at the destination. The $i$-th status update message reaches the computing queue at $t_i$ in Fig. \ref{fig:subfig:b}. In accordance with zero-wait policy, the $(i+1)$-th status update message starts to be transmitted at $t_i$. Denote $t_i^{'}$ as the time instant for terminating service of the $i$-th status update message in the MEC server.
\subsubsection{Partial Computing}
Zero-wait strategy in partial computing means that the source generates a new message when the intermediate result of the previous one is received by the remote MEC server. Thus, both the local computing queue and the transmission queue are empty. The age evolution of partial computing is shown in Fig. \ref{fig:subfig_evo:c}. The generation time instant of the $i$-th message is denoted by $g_i$, which is also the time instant $t_{i-1}$ when the $(i-1)$-th message arrives at the remote computing queue. Denote $d_i$ as the time instant when the computing of parts of the $i$-th message is finished at the local server. Denote $t_i^{'}$ as the computing completion time instant of the $i$-th message at the MEC server, at which time the age drops sharply.
\subsection{Average AoI}
Notice that local computing as well as remote computing can be viewed as special cases of partial computing. In particular, local computing can be considered as the case of partial computing with zero transmission time and zero remote computing time (or equivalently infinite transmission rate and infinite remote computing rate), and remote computing can be viewed as the special case with zero local computing time (infinite local computing rate). Thus, we firstly calculate the average AoI for partial computing. Then, the result can be easily applied to local computing and remote computing.
\subsubsection{Partial Computing}
At time $t$, a time-stamp $u(t)$ denotes the generation time of the previous processed message, and the following random process $\Delta(t)$ defines the AoI of the processed status at the target node.
\begin{equation}\label{equ:}
\Delta(t) = t-u(t).
\end{equation}
Fig. \ref{fig:subfig_evo:c} illustrates the evolution of $\Delta(t)$ with FCFS principle. At $t = 0$, the queue is empty with $ \Delta(0) = \Delta_0 $.
The average age of the processed status message is the area between the curve of $\Delta(t)$ and $t$-axis in Fig. \ref{fig:subfig_evo:c} normalized by the observed time length. The average AoI in interval $(0, \tau )$, is
\begin{equation}\label{equ:age-defi}
\Delta_\tau = \frac{1}{\tau}\int_0^{\tau}\Delta(t)dt.
\end{equation}
Set the length of the observation interval $\tau = t_{n}^{'}$. The average AoI in partial computing can be represented as
\begin{equation}\label{equ:defi_jointcom}
\Delta_\tau = \frac{\sum_{i =1}^{n}Q_i+(B_{n}+T_{n})^2/2}{\tau}.
\end{equation}
From Fig. \ref{fig:subfig_evo:c}, we know that $Q_1$ is a polygon, and $Q_i (i \geq 2)$ is an isosceles trapezoid, which can be derived from two isosceles triangles, i.e.,
\begin{equation}\label{equ:Q_jointcom}
\begin{split}
Q_i &= \frac{1}{2}(B_{i-1}+B_{i}+T_{i})^2-\frac{1}{2}(B_{i}+T_{i})^2\\
&= T_{i}B_{i-1}+B_{i}B_{i-1}+\frac{1}{2}B_{i-1}^2,
\end{split}
\end{equation}
where $B_i=g_{i+1}-g_i=t_i-t_{i-1}$ represents the inter-generation time from the $i$-th message to the $(i+1)$-th one at the source node. $B_i$ is also the time spent in local computing and transmission of the $i$-th message, i.e., $B_i = D_i+Y_i$, where $D_i=d_i-g_i$ refers to the service time of the $i$-th message in the local server and $Y_i=t_i-d_{i}$ denotes the service time in the channel. Denote $T_{i}=t_{i}^{'}-t_{i}$ as the elapsed time from the arrival time instant at the remote computing queue for the $i$-th status update message to the service termination time instant in the MEC server. A new representation of the average AoI in partial computing can be derived as
\begin{equation}\label{equ:de_inf_jointcom}
\Delta_\tau = \frac{\tilde{Q}}{\tau}+\frac{n-1}{\tau}\frac{1}{n-1}\sum_{i =2}^{n}Q_i,
\end{equation}
where $\tilde{Q}=Q_1+(B_{n}+T_{n})^2/2$. Note that the contribution of $\tilde{Q}$ to the average AoI is negligible, since when  $\tau\rightarrow \infty$, the first term in (\ref{equ:de_inf_jointcom}) divided by $\tau$ tends to zero. From Fig. \ref{fig:subfig_evo:c}, we know that $t_n^{'} = t_0+ \sum_{i=1}^n(D_i+Y_i)+T_n$, then $\lim_{\tau\rightarrow\infty}\frac{\tau}{n}=E[D_i+Y_i]$. Thus, the inverse of the fraction to the left of the second term  in (\ref{equ:de_inf_jointcom}), $\frac{\tau}{ n-1}$, can be viewed as the sum service time of the local server and the transmission channel. Thus, the following equation is obtained
\begin{equation}\label{equ:de_mu1}
\lim_{\tau\rightarrow\infty}\frac{\tau}{n}=\frac{1}{\mu_l}+\frac{1}{\mu_t},
\end{equation}
where $\mu_l$ is the local service rate and $\mu_t$ is the transmission rate.

 Combine (\ref{equ:Q_jointcom}) and (\ref{equ:de_inf_jointcom}) and let $\tau$ increase to infinity, the average AoI in partial computing can be obtained as
\begin{align}\nonumber
\bar{\Delta}_p&= \lim_{\tau\rightarrow \infty}\Delta_\tau =\frac{\mu_l\mu_t}{\mu_l+\mu_t}E[Q_i]\\\label{equ:average_1_partial}
&=\frac{\mu_l\mu_t}{\mu_l+\mu_t}\left(E[T_{i}B_{i-1}]+E[B_{i}B_{i-1}]+\frac{1}{2}E[B_{i-1}^2]\right).
\end{align}

\subsubsection{Local Computing}
Compared with partial computing, both the transmission time and the remote computing time are zero in local computing, i.e., $T_i = 0$, $B_i = D_i$. It is equivalent to infinite transmission rate and infinite remote computing rate in partial computing model, i.e., $\mu_t \rightarrow \infty$, $\mu_s \rightarrow \infty$. Therefore, the average AoI in local computing can be obtained as follows
\begin{align}\nonumber
\bar{\Delta}_l&=\bar{\Delta}_p\Big|_{\mu_t\rightarrow\infty,\mu_s\rightarrow\infty}\\
&=\!\!\frac{\mu_l\mu_t}{\mu_l+\!\mu_t}\!\left(E[T_{i}B_{i-1}]\!+E[B_{i}B_{i-1}]\!+\!\frac{1}{2}E[B_{i-1}^2]\!\right)\Big|_{\mu_t\!\rightarrow\infty,\mu_s\!\rightarrow\infty}\\\label{equ:average_local}
&=\mu_l\left(E[D_{i}D_{i-1}]+\frac{1}{2}E[D_{i-1}^2]\right).
\end{align}

\subsubsection{Remote Computing}
Compared with partial computing, the local computing time in remote computing is zero, i.e., $B_i = Y_i$. It is equivalent to infinite local computing rate in partial computing model, i.e., $\mu_l \rightarrow \infty$. Hence, the average AoI in remote computing can be obtained as follows
\begin{align}\nonumber
\bar{\Delta}_r&=\bar{\Delta}_p\Big|_{\mu_l\rightarrow\infty}\\
&=\!\!\frac{\mu_l\mu_t}{\mu_l+\mu_t}\!\!\left(E[T_{i}B_{i-1}]\!+E[B_{i}B_{i-1}]\!+\frac{1}{2}E[B_{i-1}^2]\right)\Big|_{\mu_l\rightarrow\infty}\\\label{equ:average_1_remote}
&=\mu_t\left(E[T_{i}Y_{i-1}]+E[Y_{i}Y_{i-1}]+\frac{1}{2}E[Y_{i-1}^2]\right).
\end{align}

Noted that for tandem queue, in the case where one or the other queue is unstable, the whole system will be unstable and the AoI will reach infinity. In this paper, as we adopt zero-wait policy in the first hop, the stability issue of the tandem queue only exists in the second hop. Therefore, as long as the service rate of the first hop is less than that of the second hop, the tandem queue is stable. Hence, the stability issues for the three cases can be discussed as follows. For local computing, the queue is stable due to the infinitely large transmission rate. For remote computing, if the remote computing rate is equal or less than the transmission rate, the tandem queue will be unstable. For partial computing, when the remote computing rate is equal or less than $\frac{\mu_l\mu_t}{\mu_l+\mu_t}$, the whole system will be unstable.

The calculation of equation (11) depends on the distributions of transmission time and computing time. In this paper, we assume exponentially distributed transmission time to indicate purely random transmission process, and derive the results for two different computing time distributions: exponential distribution \cite{Sthapit2019Computational} and deterministic distribution. The main results are detailed in the following sections. Noted that the main results are obtained based on the whole system to be stable.
\section{Average AoI with Exponentially Distributed Computing Time}
We will deduce the average age with exponentially distributed computing time in this section. Exponential distribution is widely used to model random events in practice and can derive closed-form analytical results in most cases. For example, security based on face recognition identifies the user as a random event, and users' arrival process usually can be viewed as a Poisson process. In this section, closed-form average AoIs in the three computing schemes are presented.
\subsection{Local Computing}
The average AoI in local computing with exponentially distributed computing time is defined as below.
\begin{theorem}
If the local computing time is exponentially distributed, the average AoI of local computing (\ref{equ:average_local}) is expressed as
\begin{equation}\label{equ:average_main_local}
\bar{\Delta}_l= \frac{2}{\mu_l},
\end{equation}
where $\mu_l$ is the computing rate of the local server.
\end{theorem}

\begin{IEEEproof}
See Appendix A.
\end{IEEEproof}

\subsection{Remote Computing}
We obtain the closed-form expression of average AoI for remote computing using zero-wait policy. Both the transmission time of the channel and the computing time at the MEC server are exponentially distributed.
\begin{theorem}
Assume both the transmission time and the remote computing time are exponentially distributed. The average AoI in remote computing (\ref{equ:average_1_remote}) is expressed as
\begin{equation}\label{equ:average_main_remote}
\bar{\Delta}_r = \frac{1}{\mu_s}\left(\frac{2\mu_t^3-\mu_t^2\mu_s+\mu_t\mu_s^2}{\mu_s(\mu_s+\mu_t)(\mu_s-\mu_t)}+\frac{2\mu_s}{\mu_t}+1\right),
\end{equation}
where $\mu_t$ denotes the transmission rate and $\mu_s$ is the computing rate of the MEC server.
\end{theorem}

\begin{IEEEproof}
See Appendix B.
\end{IEEEproof}

\subsection{Partial Computing}
Now we derive the closed-form average AoI for partial computing under zero-wait policy. The computing time of the local server and the MEC server, as well as the transmission time of the channel are exponentially distributed.
\begin{theorem}
Assume the local computing time, the channel transmission time and the remote computing time are exponentially distributed. The average AoI in partial computing (\ref{equ:average_1_partial}) is obtained as
\begin{align}\nonumber
\bar{\Delta}_p&=\frac{1}{\mu_s}+\frac{\phi-1}{\mu_l}-\frac{1+\varphi}{\mu_t}+\frac{\mu_l\mu_t}{\mu_l+\mu_t}\left(\frac{1}{\mu_l\mu_s}+\frac{1}{\mu_t\mu_s} +\frac{2}{\mu_l^2}\right.\\\nonumber
&\left.+\frac{2}{\mu_t^2}+\frac{3}{\mu_l\mu_t}+\frac{\mu_l\mu_t}{\mu_t-\mu_l}\left(\frac{1}{(\xi+\mu_l)^2}-\frac{1}{(\xi+\mu_t)^2}\right)\right.\\\label{equ:main_joint}
&\left.\left(\frac{1}{\xi}- \frac{\phi}{(\mu_l+\xi)}+\frac{\varphi}{(\mu_t+\xi)}\right)\right),
\end{align}
where $\phi=\frac{\mu_t\mu_s}{(\mu_t-\mu_l)(\mu_l+\mu_s)}$ and $\varphi=\frac{\mu_l\mu_s}{(\mu_t-\mu_l)(\mu_t+\mu_s)}$. The notation $\xi$ is expressed as
\begin{equation}\label{equ:xi}
\xi=\frac{\mu_s-(\mu_l+\mu_t)+\sqrt{(\mu_s-\mu_t+\mu_l)^2+4\mu_t\mu_s}}{2}.
\end{equation}
\end{theorem}
\begin{IEEEproof}
See Appendix C.
\end{IEEEproof}
It is remarkable that the average AoI for both local computing (\ref{equ:average_main_local}) and remote computing (\ref{equ:average_main_remote}) can be obtained based on (\ref{equ:main_joint}). Since partial computing with infinite transmission rate and infinite remote computing rate can be taken as local computing, while partial computing with infinite local computing rate can be viewed as remote computing. Thus, we have
\begin{equation}\label{equ:average_main_local_equa}
\bar{\Delta}_l=\bar{\Delta}_p\Big|_{\mu_t\rightarrow\infty,\mu_s\rightarrow\infty} =\frac{2}{\mu_l},
\end{equation}
\begin{equation}\label{equ:average_main_remote_equa}
\bar{\Delta}_r\!=\!\bar{\Delta}_p\Big|_{\mu_l\rightarrow\infty} \!=\!\frac{1}{\mu_s}\left(\frac{2\mu_t^3-\mu_t^2\mu_s+\mu_t\mu_s^2}{\mu_s(\mu_s+\mu_t)(\mu_s-\mu_t)}+\frac{2\mu_s}{\mu_t}+1\right).
\end{equation}

\begin{figure}
\centering
    \includegraphics[width=2.7in,height = 2.4in]{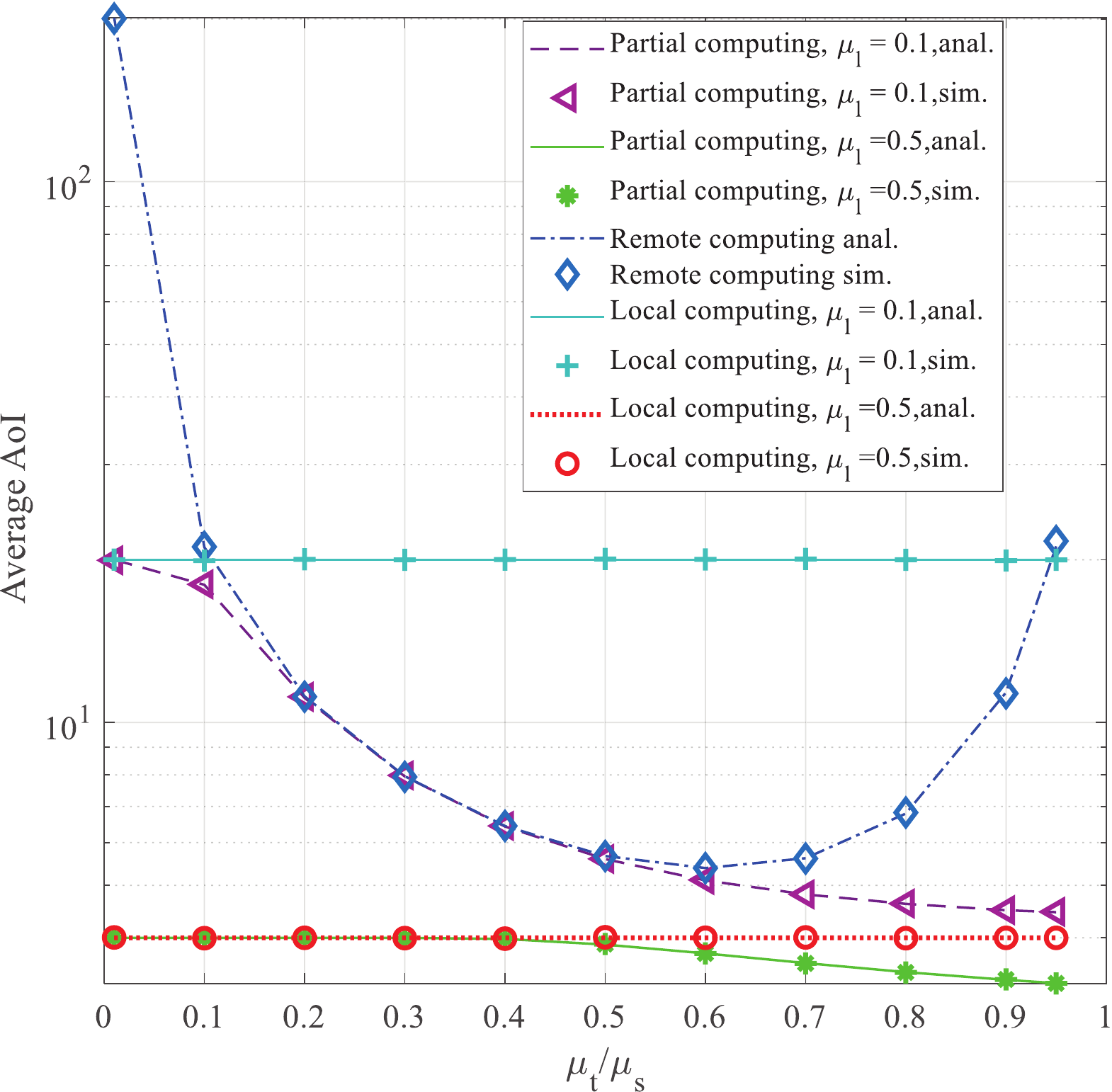}
\caption{Comparison of the average AoI among the three computing schemes with exponentially distributed computing time under different $\mu_t/\mu_s$. The remote computing rate is assumed as $\mu_s = 1$.} \label{fig:joint_remote}
\end{figure}

As shown in Fig. \ref{fig:joint_remote}, the analytical results of the three computing schemes with exponentially distributed computing time  are validated by
simulations. It must be clear that both local computing and remote computing share the same settings of $\mu_l$, $\mu_t$ and $\mu_s$. While for partial computing, we adopt a linear computation partitioning model \cite{Wang2016Mobile}. In particular, the computing rate of the local server, the transmission rate and the computing rate of the MEC server are denoted as $\mu_l/(1-\alpha)$, $\mu_t/\alpha$, $\mu_s/\alpha$, respectively, and $\alpha\in[0,1]$ denotes the percentage of computing tasks computed by the MEC server. Particularly, $\alpha = 0$ in local computing while $\alpha = 1$ in remote computing.
In Fig. \ref{fig:joint_remote}, we set $\mu_s = 1$ and denote $\rho_s=\mu_t/\mu_s$, the optimal value of $\alpha$ in partial computing is determined with numerical simulation methods according to the criteria of minimizing the average AoI in partial computing.
In this figure, the average AoI in partial computing is smaller than local and remote computing. And when $\rho_s<0.1$ or $\rho_s>0.9$, partial computing is significantly better than remote computing. It can be explained as when $\rho_s$ is small, outdated message will occur due to the long transmission time in remote computing, and when $\rho_s$ is large, the MEC server will compute delayed message since the status update messages are backlogged in the computing queue. While for partial computing, it can execute a proper proportion of computing tasks required by a computation-intensive message locally to mitigate the above problems. The performance of local computing is the same as partial computing when the $\rho_s$ is small. The reason is that when the transmission rate is sufficiently small, partial computing prefers to compute the messages locally. As $\rho_s$ increases, the average AoI in partial computing is smaller than that in local computing, especially for a small local computing rate, such as $\mu_l = 0.1$.
\section{ Average AoI with Deterministic Computing Time}
In this part, we derive the average AoI with deterministic computing time. In the application scenario where the volume of a computing task is constant and the computing resource allocated to the task is static, the computing time for an update message is deterministic.
The analysis of the average AoI in the three computing schemes with deterministic computing time is provided in the following.
\subsection{Local Computing}
The average AoI in local computing with deterministic computing time is given in the following theorem.
\begin{theorem}
If the local computing time is deterministic and is equal to $1/\mu_l$, the average AoI for local computing (\ref{equ:average_local}) is expressed as
\begin{equation}\label{equ:average_main_local_deter}
\bar{\Delta}_l= \frac{3}{2\mu_l}.
\end{equation}
\end{theorem}

\begin{IEEEproof}
See Appendix D.
\end{IEEEproof}
\subsection{Remote Computing}
With deterministic computing time, the closed-form expression of average AoI for remote computing is difficult to obtain. But the result can be numerically calculated based on the following theorem.
\begin{theorem}
Assume the transmission time of the channel is exponentially distributed with mean $1/\mu_t$ and the computing time at the MEC server is deterministic and is equal to $1/\mu_s$.
The average AoI can be numerically computed as:
\begin{align}
\bar{\Delta}_r\!=\mu_t\!\left(\int_{0}^{\infty}y_{i}E[W_{i}|Y_{i-1}=y_i]f_{Y}(y_i)~dy_i\!+\!\frac{1}{\mu_t\mu_s}\!+\!\frac{2}{\mu_t^2}\right),
\end{align}
where
\begin{align}\label{equ:condi_solu_det_upda}\nonumber
&E[W_{i}|Y_{i-1} = y_i]
\!=\!\int_{0}^{y_i-1/\mu_s}\!\!f_{W}(w)\!\!\int_{0}^{1/\mu_s}\!\!\!f_{Y}(y)(1/\mu_s\!-\!y)\!~dydw \\ &+\!\!\!\int_{y_i-1/\mu_s}^{\infty}\!\!\!f_{W}(w)\!\!\!\int_{0}^{w-\!y_i+2/\mu_s}\!\!\!f_{Y}(y)(w\!-\!y_i+2/\mu_s\!-y\!)\!~dydw,
\end{align}
where probability density function $f_{Y}(y)=\mu_te^{-\mu_ty}$ and $f_{W}(w)$ can be obtained as the derivative of the cumulative distribution function (CDF)
\begin{align}\label{equ:wait_det}
F_W(w) = (1-\rho)\sum_{k=0}^{\lfloor w\mu_s\rfloor}\frac{\rho^k}{k!}(k- w\mu_s)^ke^{\rho(w\mu_s-k)},
\end{align}
where $\rho=\mu_t/\mu_s$.
\end{theorem}
\begin{IEEEproof}
See Appendix E.
\end{IEEEproof}

\subsection{Partial Computing}For partial computing, if we view the local computing process and the transmission process as a whole, the remote computing part can be considered as an $GI/D/1$ queuing model. If $\mu_s \geq \mu_l$, a closed-form expression can be obtained. Otherwise, the average AoI can only be calculated numerically. The result is concluded in the theorem below.
\begin{theorem}
Assume both the local computing time $1/\mu_l$ and the remote computing time $1/\mu_s$ are deterministic, and the channel transmission time is exponentially distributed with mean $1/\mu_t$. If $\mu_s\geq \mu_l$, the average AoI in partial computing is expressed as
\begin{equation}\label{}
 \bar{\Delta}_p=\frac{1}{\mu_l+\mu_t}\left(3+\frac{3\mu_t}{2\mu_l}+\frac{2\mu_l}{\mu_t}+\frac{\mu_t}{\mu_s}+\frac{\mu_l}{\mu_s}\right).
\end{equation}
For $\mu_s< \mu_l$, the average AoI can be numerically computed as follows:
\begin{align}\nonumber
\bar{\Delta}_p
&=\frac{\mu_l\mu_t}{\mu_l+\mu_t}\left(\int_{\frac{1}{\mu_l}}^{\infty}E[W_{i}|B_{i-1}=b_i]b_{i}f_B(b_i)~db_i+\frac{1}{\mu_s\mu_l}\right.\\
&\left.+\frac{1}{\mu_s\mu_t}+\frac{3}{\mu_l\mu_t}+\frac{2}{\mu_l^2}\!+\frac{2}{\mu_t^2}\right),
\end{align}
where
\begin{align}\nonumber
&E[W_{i}|B_{i-1} \!= b_i]
\!=\!\!\int_{0}^{b_i-1/\mu_s}\!\!f_{W}(w)\!\int_{\frac{1}{\mu_l}}^{\frac{1}{\mu_s}}\!\!f_{B}(b)(1/\mu_s\!-\!b)\!\!~dbdw\\ &+\int_{b_i\!-1/\mu_s}^{\infty}\!\!\!\!f_{W}(w)\!\!\!\int_{\frac{1}{\mu_l}}^{w\!-\!b_i\!+\frac{2}{\mu_s}}\!\!\!\!f_{B}(b)(w\!-\!b_i+2/\mu_s\!-\!b)\!\!~dbdw, \label{equ:condi_solu_jointcom_det}
\end{align}
where probability density function $f_{B}(b)=\mu_te^{-\mu_t(b-\frac{1}{\mu_l})}$ and $f_{W}(w)$ is the derivative of the CDF
\begin{align}\label{equ:wait_det}
F_W(w) = (1-\rho)\sum_{k=0}^{\lfloor w \mu\rfloor}\frac{\rho^k}{k!}(k- w\mu)^ke^{\rho(w\mu-k)},
\end{align}
where $\mu = \frac{\mu_l\mu_s}{\mu_l-\mu_s}$, and $\rho=\mu_t/\mu$.
\end{theorem}
\begin{IEEEproof}
See Appendix F.
\end{IEEEproof}

As shown in Fig. \ref{fig:three_D}, the analytical results of the three computing schemes with deterministic computing time are validated by simulations, and the remote computing rate $\mu_s=1$.
When $\mu_l = $ 0.5, local computing and partial computing perform similar for small and large $\mu_t/\mu_s$. The reason is that when the transmission rate $\mu_t$ is small, the transmission time is very large; while the transmission rate is large, the computing queue at the MEC server is long. Thus, in the two cases, partial computing will process the whole computation-intensive message at the local server. While for medium value of $\mu_t/\mu_s$, partial computing attains smaller average AoI compared with local computing due to the benefit of splitting the computing task between local and remote servers.
The average AoI in remote computing is always larger than partial computing, and is even larger than local computing when $\mu_l = $ 0.5. It shows that there is no benefit of using remote computing when the local computing capacity is sufficient.
\begin{figure}
\centering
    \includegraphics[width=2.7in,height =2.4in]{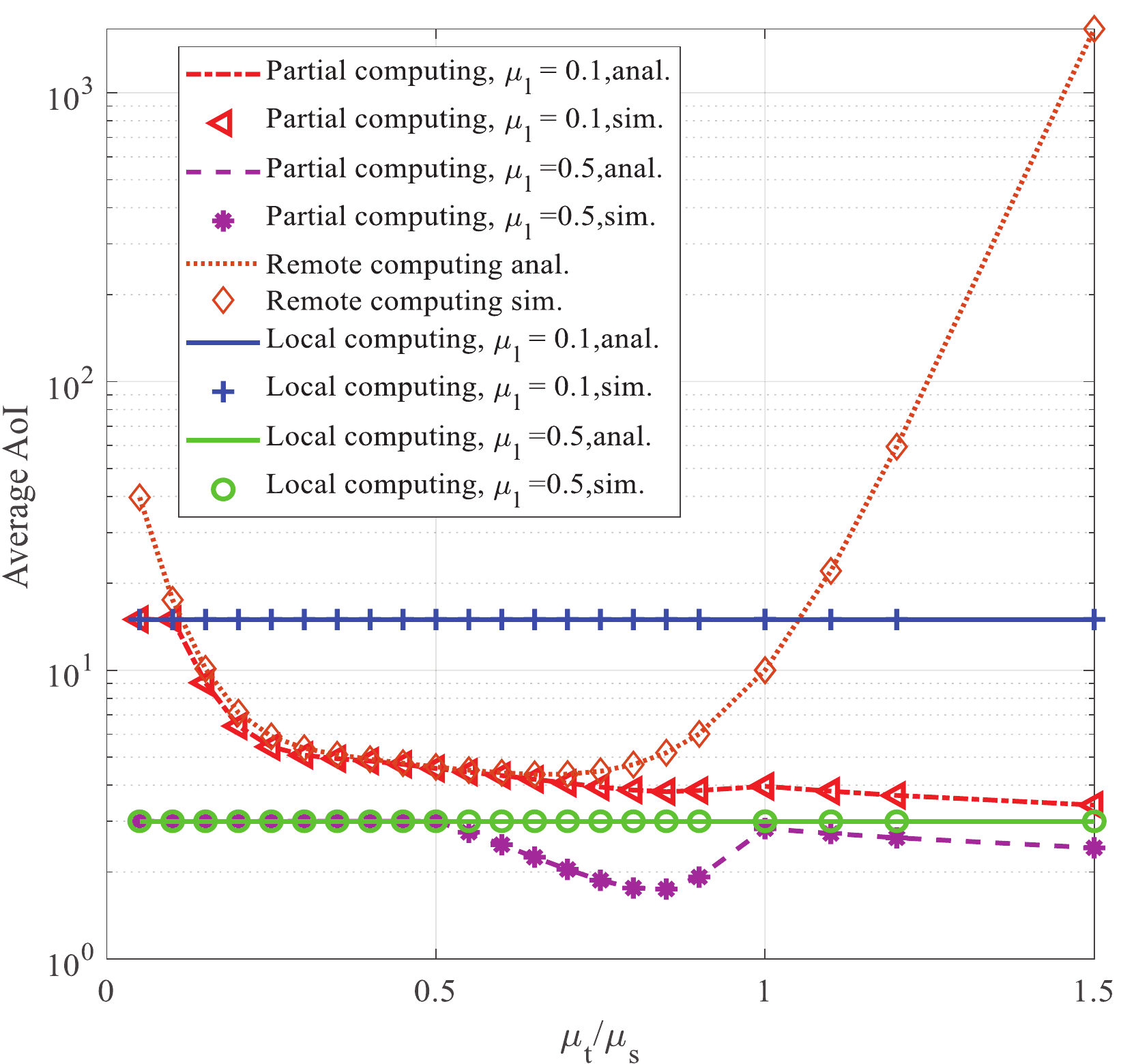}
\caption{Comparison of the average AoI among the three computing schemes with deterministic computing time under different $\mu_t/\mu_s$. The remote computing rate is assumed as $\mu_s = 1$.} \label{fig:three_D}
\end{figure}
\section{Numerical Analysis}
Different parameters will influence the average AoI for computation-intensive messages in MEC systems.
In this section, we study with the exponentially distributed computing time on the influence of various parameters, including message size, required number of CPU cycles, average data rate and computing capacity of the MEC server. Note that, we show the numerical results based on the stable system.
For all status update messages, we adopt identical parameter pair $(l,c)$ to describe the status update message, where $l$ is the input size of the message and $c$ indicates the required number of CPU cycles to compute the original message.
The size of the transmitted data and the data rate will affect the transmission time. Note that, the transmission time refers to the total time for delivering a message over multiple channel uses. Due to channel fading, the number of bits that can be transmitted in one channel use is random. Therefore, it is reasonable to assume random transmission time for each message.
 The required number of CPU cycles and the computing capacity will affect the computing time. Let $f_1$ be the average local computing capacity and denote $f_s$ as the average available computing capacity of the MEC server. Denote $R$ as the average data rate of the channel. We adopt a linear model to represent their relationships, then the service rates $\mu_l$, $\mu_t$, $\mu_s$ can be expressed as \cite{Wang2016Mobile}
 \begin{equation}\label{equ:mul}
\mu_{l} = f(c,f_l,\alpha) = \frac{f_l}{(1-\alpha)c},
\end{equation}
 \begin{equation}\label{equ:mu4}
\mu_{t} = f(l,R,\alpha) = \frac{R}{\alpha l},
\end{equation}
 \begin{equation}\label{equ:mus}
\mu_{s} = f(c,f_s,\alpha) = \frac{f_s}{\alpha c}.
\end{equation}
Recall that $\alpha\in[0,1]$ denotes the percentage of computing tasks computed by the MEC server. For local computing, $\alpha=0$, while for remote computing $\alpha=1$, and $\alpha\in(0,1)$ represents partial computing scheme. For partial computing, given other parameters, $\alpha$ is chosen to minimize the average AoI.

\begin{figure}
\centering
    \includegraphics[width=2.7in,height =2.4in]{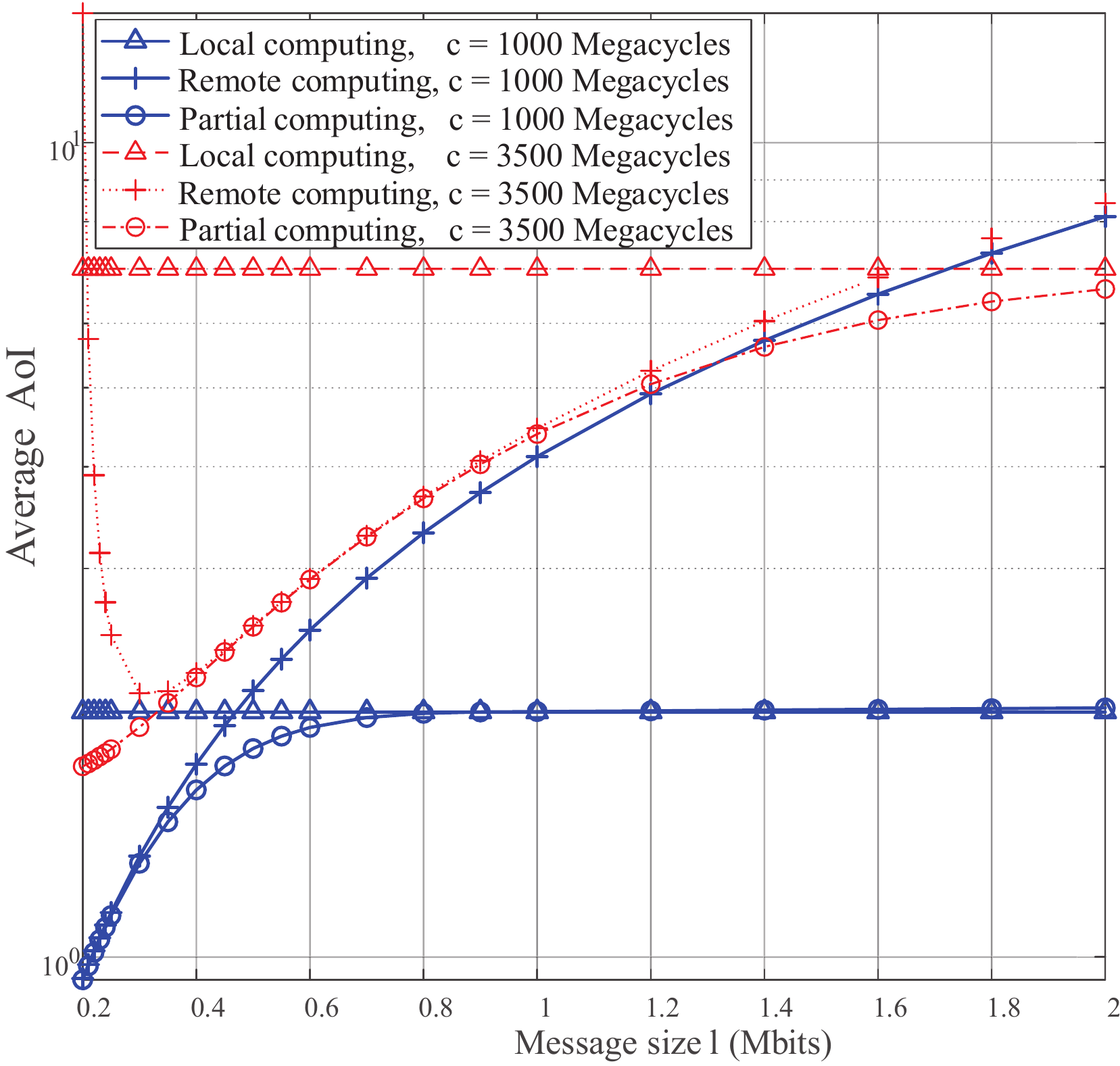}
\caption{Average AoI versus message size} \label{fig:message}
\end{figure}

Firstly, set $R = 0.5$ Mbits/s, $f_l = 1$ GHz and $f_s = 9$ GHz.
The average AoI associated with the message size $l$ with different required number of CPU cycles $c$ is shown in Fig. \ref{fig:message}.
 It is shown that the AoI for local computing keeps stable when the message size is increasing.
For remote computing, the average AoI firstly decreases sharply and then increases gradually as the message size increases with c = 3500 Megacycles. This is because the transmission rate is large with small message size, which makes a large amount of messages queued in the computing queue waiting to be computed, while for the large message size, the transmission time takes longer.
When $c= 1000$ Megacycles, there is a cross point between the curves for local computing and remote computing at $l\approx 0.47$ Mbits; meanwhile the cross point is at $l\approx 1.64$ Mbits when $c=3500$ Megacycles. This phenomenon means whether remote computing is superior to local computing or not depends on the message size.
 With the decrease of message size, the average AoI for partial computing decreases. By properly partitioning the computing tasks between the local server and the remote server, the performance of partial computing is always better than the other two schemes with the same parameters setting. Moreover, with sufficiently large message size, the average AoI for partial computing is the same as local computing. This is because the larger message size will cause the longer transmission time. With the aim to reduce the average AoI, the optimal parameter in partial computing is $\alpha=0$, which is equivalent to local computing.

\begin{figure}
\centering
    \includegraphics[width=3in,height = 2.4in]{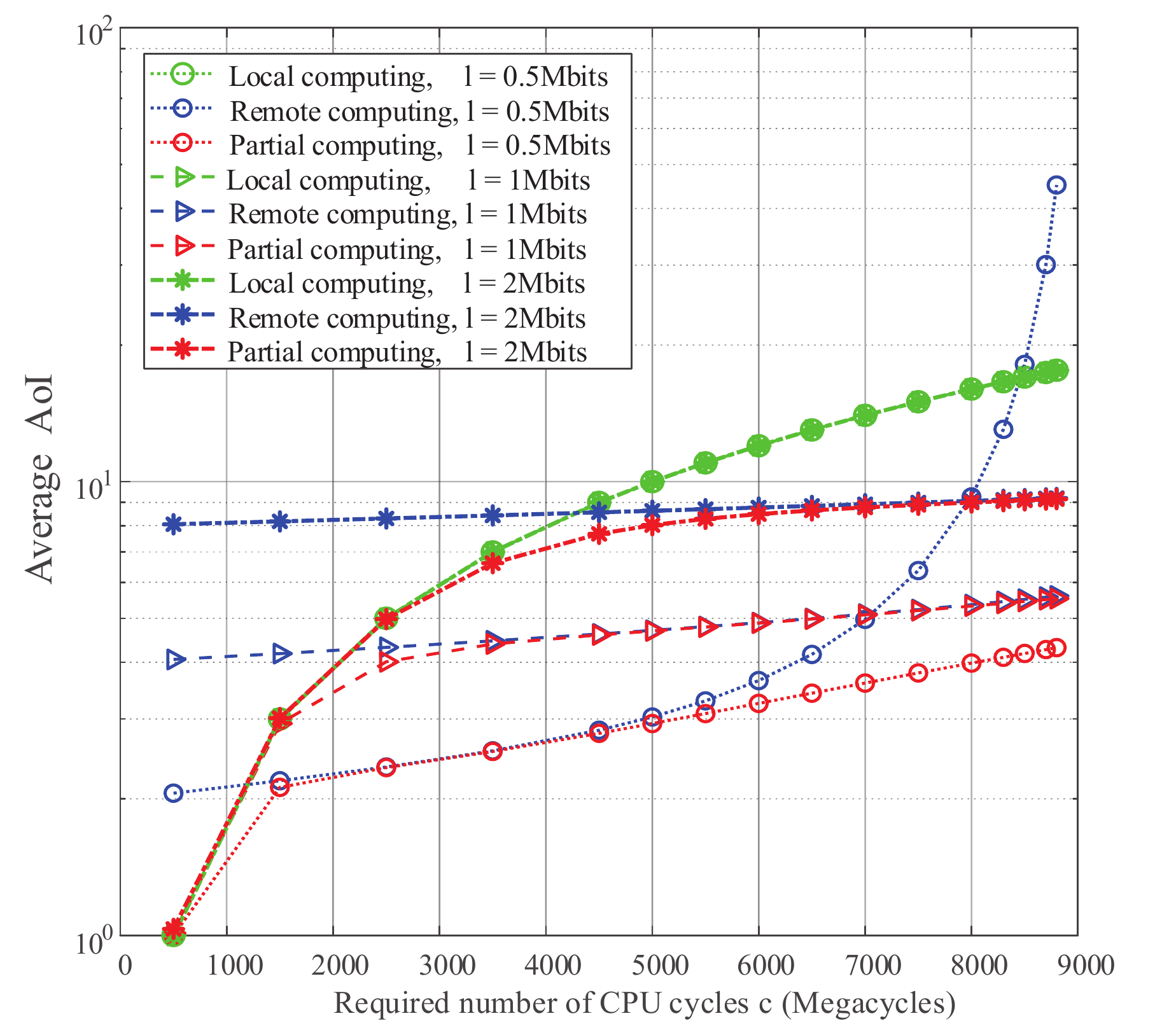}
\caption{Average AoI versus required number of CPU cycles} \label{fig:CPU}
\end{figure}

Fig. \ref{fig:CPU} depicts the average AoI versus the required number of CPU cycles $c$ with different message sizes, where $R = 0.5$ Mbits/s, $f_l = 1$ GHz and $f_s = 9$ GHz. It can be seen that the AoI curves of the three schemes rise up when the required number of CPU cycles is increasing, due to the increased computation time.
There is an overlap among the curves for local computing with different message sizes because the average AoI of local computing is not affected by the message size. In general, remote computing outperforms local computing for large number of CPU cycles. However, under the condition of $c\geq 7000$ Megacycles, the average AoI of remote computing with $l=0.5$ Mbits boosts and performs worse than local computing. This is because as $c$ increases, the MEC server computes slowly which results in long queuing delay, and hence leads to large AoI. For $l=1$ Mbits or $l=2$ Mbits, a sudden sharp increase of the average AoI in remote computing occurs on the larger required number of CPU cycles for the same reason.
\begin{figure}
\centering
    \includegraphics[width=2.7in,height = 2.4in]{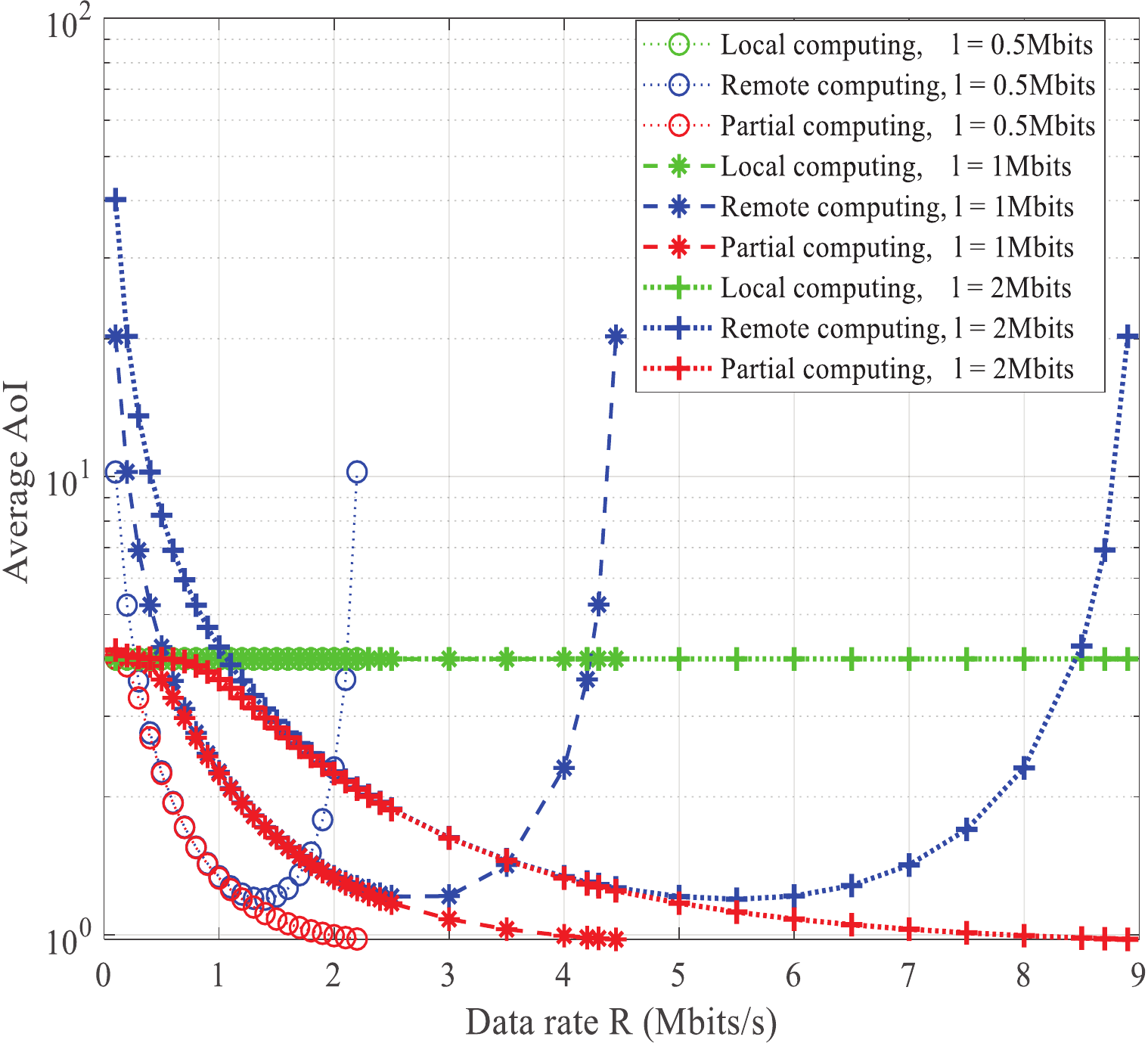}
\caption{Average AoI versus data rate} \label{fig:data}
\end{figure}

In Fig. \ref{fig:data}, we set $c=2000$ Megacycles, $f_l = 1$ GHz and $f_s = 9$ GHz and show the AoI performance versus data rate $R$.
Firstly, for remote computing with $l=0.5$ Mbits, the average AoI firstly decreases with the increasing of data rate, since higher data rate leads to less transmission time.
However, the average AoI increases when $R\geq 1.4$  Mbits/s. This is because the number of messages queuing at the MEC server also increases as data rate increases.
For remote computing  with $l=1$ Mbits and $l=2$ Mbits, the average AoI has the same trend as $l=0.5$  Mbits and it increases after a certain point of larger data rate.
Secondly, for partial computing, the average AoI always decreases as data rate increases. Because the transmission time reduces as data rate increases and the congestion of queuing messages at the MEC server can be alleviated by completing a part of the computing tasks locally.
Thirdly, when the data rate is small or large, the performance of local computing is superior to remote computing. This means when data rate is small or large, there's no benefit in transmitting to the MEC server for reducing the average AoI of computation-intensive messages.
Finally, regardless of the data rate or the message size, partial computing always derives a smaller average AoI than the other two computing schemes. Thus it proves that partial computing outperforms both local computing and remote computing.

\begin{figure}
\centering
    \includegraphics[width=3in,height = 2.4in]{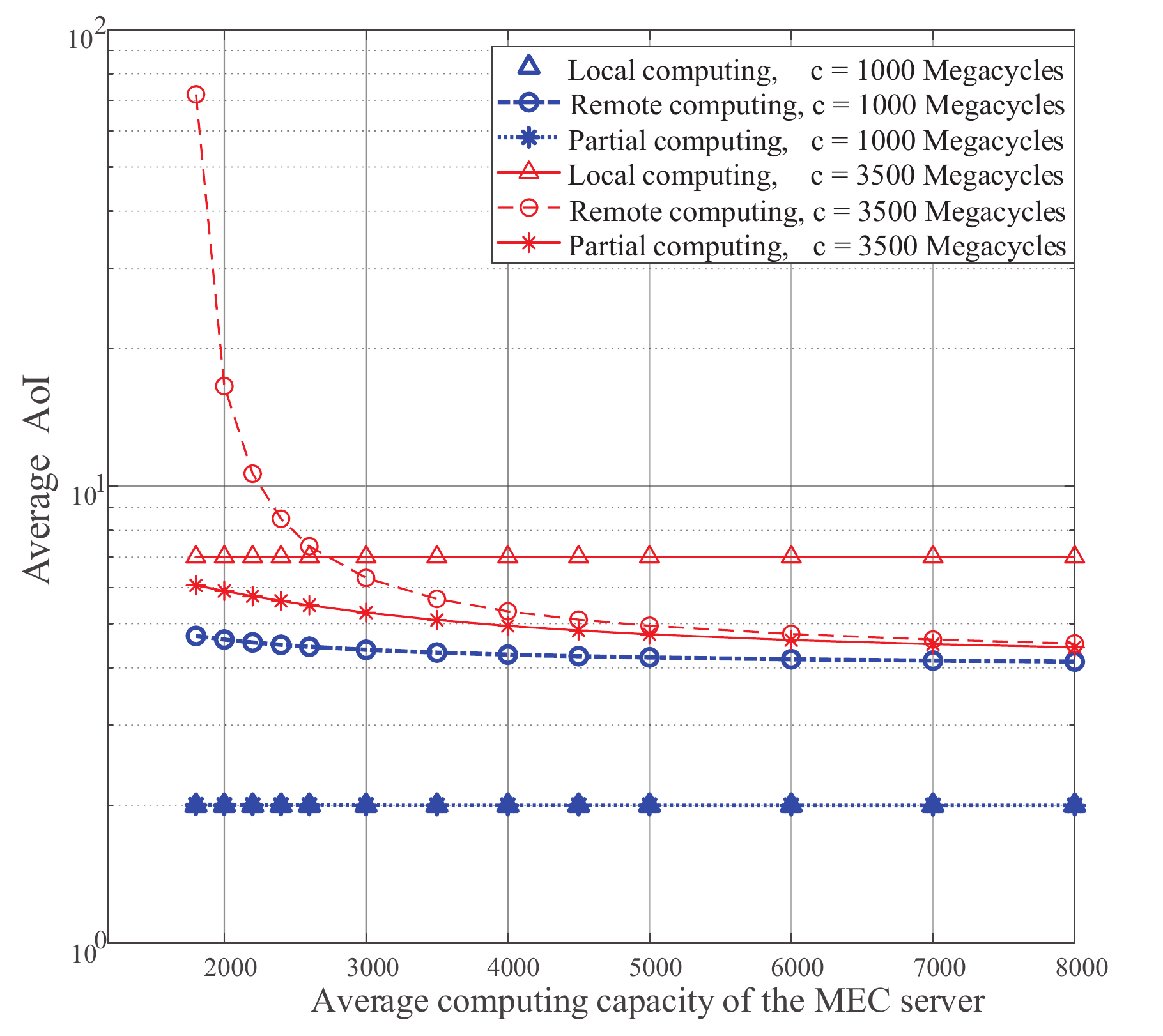}
\caption{Average AoI versus average computing capacity of the MEC server} \label{fig:computing}
\end{figure}
Fig. \ref{fig:computing} shows the impact of the computing capacity of the MEC server with data rate $R=0.5$ Mbits/s, local computing capacity $f_l = 1$ GHz and message size $l = 1$ Mbits. When $c=1000$ Megacycles, remote computing performs worse than local computing.
This is due to the required number of CPU cycles is small, the local server can complete the computation in a short time.  Under the condition of large number of required CPU cycles, such as $c=3500$ Megacycles, if the computing capacity is smaller than 2700, the average AoI of remote computing is larger than that of local computing, and vice versa. It eventually converges to the minimum average age $\bar{\Delta}_{min}=\frac{2}{\mu_t}$.
When $c=1000$ Megacycles, the curves for local computing and partial computing are overlapped, which indicates that local computing performs the best. When $c=3500$ Megacycles, with the increasing of the average computing capacity of the MEC server, the performance of partial computing converges to remote computing. Thus, it is better to offload most of the computing task to the MEC server.

\section{CONCLUSIONS}
We have analyzed the AoI for computation-intensive messages in an MEC system with three computing strategies: local computing, remote computing and partial computing. Two computing time distributions are considered: exponential distribution and deterministic distribution. The closed-form expressions for the three computing schemes with exponentially distributed computing time are derived.
And the average AoI with deterministic computing time is obtained numerically.
 Simulation results prove that partial computing has smaller average AoI than the other two computing schemes in most cases,
  and only in some special cases has the same performance as the other two schemes.
The influence of various parameters for data processing on the average AoI for exponentially distributed computing time is studied by numerical results, including message size, required number of CPU cycles, data rate, and average computing capacity of the MEC server.
We find that for computation-intensive data, the combination with MEC significantly helps to obtain the optimal AoI. For the future works, we can extend to multi source-destination pairs and consider other message generation policies.

\appendices
\section{Proof of Theorem 1}
Because the local computing time $D_i$ are independent identically distributed (iid) exponentials with $E[D_i] = 1/\mu_l$. As $D_{i-1}$ and $D_i$ are independent, we have
\begin{equation}\label{equ:exp_d}
E[D_{i}D_{i-1}]=(E[D_i])^2 =1/\mu_l^2,
\end{equation}
\begin{equation}\label{equ:exp_d_sq}
 E[D_{i-1}^2]=2/\mu_l^2.
\end{equation}
Submit (\ref{equ:exp_d}) and (\ref{equ:exp_d_sq}) into (\ref{equ:average_local}), the average AoI for local computing can be obtained as
\begin{equation}\label{equ:average_local_age}
\bar{\Delta}= \lim_{\tau\rightarrow \infty}\Delta_\tau =\mu_lE[Q_i]=\frac{2}{\mu_l}.
\end{equation}
Thus, Theorem 1 can be proven.
\section{Proof of Theorem 2}
Since the arrival process of the computing queue equals to the message's departure process of the transmission channel. And the process is a Poisson process due to the zero-wait policy. Thus, the computing queue and an MEC server form an $M/M/1$ system. Consequently, the inter-arrival time $Y_i$ as well as the service time is iid exponential with $E[Y_i] = 1/\mu_t$ and average service time $1/\mu_s$. As $Y_{i-1}$ and $Y_i$ are independent, we have
\begin{equation}\label{equ:exp_y}
E[Y_{i}Y_{i-1}]=(E[Y_i])^2 =1/\mu_t^2,
\end{equation}
\begin{equation}\label{equ:exp_y_sq}
 E[Y_{i-1}^2]=2/\mu_t^2.
\end{equation}
Then we thoroughly calculate $E[T_{i}Y_{i-1}]$. For $T_i$ of status update $i$, it also shows the system time in queuing theory, which has two parts as waiting time and service time, i.e.,
\begin{equation}\label{equ:sys_time}
T_{i} = W_i+S_i,
\end{equation}
where $W_i$ is the waiting time in the computing queue and $S_i$ is the service time at the MEC server. The waiting time $W_i$ has a relation with the system time of the $(i-1)$-th message, $T_{i-1}$, and the inter-arrival time $Y_i$. Particularly,
if $T_{i-1}>Y_i$, i.e., message $i$ will reach the computing queue, while the $(i-1)$-th message is still waiting in the queue or under service, we have $W_i = T_{i-1}-Y_i$. Otherwise, $W_i=0$. Thus, we can express the waiting time of message $i$ as
\begin{equation}\label{equ:wait_time}
W_i =(T_{i-1}-Y_i)^+.
\end{equation}
From (\ref{equ:sys_time}), the term $E[T_{i}Y_{i-1}]$ can be written as
\begin{equation}\label{equ:ex_1_remote}
\begin{split}
E[T_{i}Y_{i-1}]&=E[(W_{i}+S_{i})Y_{i-1}]\\
&=E[W_{i}Y_{i-1}]+E[S_{i}Y_{i-1}].
\end{split}
\end{equation}
According to (\ref{equ:sys_time}) and (\ref{equ:wait_time}), we can obtain the term $W_{i}$,
\begin{equation}\label{equ:}
\begin{split}
W_{i}& = (T_{i-1}-Y_{i})^+=(W_{i-1}+S_{i-1}-Y_{i})^+\\
 &= ((T_{i-2}-Y_{i-1})^{+}+S_{i-1}-Y_{i})^+.
 \end{split}
\end{equation}
Notice that the system time $T_{i-2}$ relies on the service time and the waiting time of message $(i-2)$, hence it is independent of $S_{i-1}$, $Y_{i}$ and $Y_{i-1}$. Further more, the system time $T_{i}$ becomes stochastically identical, i.e., $T=^{st}T_i=^{st}T_{i-1}=^{st}T_{i-2}$, since the system will reach a stable state.
The probability density function of the system time $T$ for the $M/M/1$ system is \cite{Papoulis2001Probability}
\begin{equation}\label{equ:}
f_{T}(t) = (\mu_s-\mu_t)e^{-(\mu_s-\mu_t)t},~~~ t\geq 0.
\end{equation}
The condition expected waiting time $W_{i}$ given $Y_{i-1} = y_i$ can be derived as
\begin{align}\nonumber
&E[W_{i}|Y_{i-1} = y_i]=E[((T_{i-2}-y_{i})^{+}+S_{i-1}-Y_{i})^+|Y_{i-1} = y_i]\\\nonumber
=&E[((T_{i-2}-y_{i})^{+}+S_{i-1}-Y_{i})^+]\\\nonumber
=&\int_{0}^{\infty}f_{T}(t)\int_{0}^{\infty}f_{S}(s)\int_{0}^{\infty}f_{Y}(y)\left((t-y_i)^++s-y\right)^+dy dsdt,\\\nonumber
=&\int_{0}^{y_i}f_{T}(t)\int_{0}^{\infty}f_{S}(s)\int_{0}^{s}f_{Y}(y)(s-y)~dydsdt \\\nonumber \label{equ:condi_solu} &+\!\int_{y_i}^{\infty}\!f_{T}(t)\int_{0}^{\infty}\!f_{S}(s)\int_{0}^{t-y_i+s}\!f_{Y}(y)(t-y_i+s-y)~dydsdt,\\
=&\frac{2\mu_t}{(\mu_s+\mu_t)(\mu_s-\mu_t)}e^{-(\mu_s -\mu_t)y_i}+\frac{\mu_t}{\mu_s(\mu_s+\mu_t)}.
\end{align}

Returning to (\ref{equ:ex_1_remote}), we note that the inter-arrival time $Y_{i-1}$ is independent of $S_{i}$, the service time of the MEC server for the $i$-th message. Thus, equation (\ref{equ:ex_1_remote}) can be represented as
\begin{equation}\label{equ:ex_1_ty_remote}
E[T_{i}Y_{i-1}] =E[W_{i}Y_{i-1}]+E[S_{i}]E[Y_{i-1}],
\end{equation}
where $E[S_{i}]=\frac{1}{\mu_s}$. Further more, adopting the conditional expectation in (\ref{equ:condi_solu}), we can have equation (\ref{equ_22})$-$(\ref{equ:wy_remote}).
\begin{align}
\label{equ_22}&E[W_{i}Y_{i-1}]=\int_{0}^{\infty}y_{i}E[W_{i}|Y_{i-1}=y_i]f_{Y_i}(y_i)~dy_i\\
\label{equ_23}&=\!\!\int_{0}^{\infty}\!\!\!\!y_{i}\!\left(\frac{2\mu_t}{(\mu_s\!+\mu_t)(\mu_s\!-\mu_t\!)}e^{-(\mu_s -\mu_t)y_i}\!+\!\!\frac{\mu_t}{\mu_s(\mu_s\!+\!\mu_t)}\right)\!\mu_te^{-\mu_ty_i}~\!\!dy_i\\
\label{equ:wy_remote}&=\frac{2\mu_t^2+\mu_s^2-\mu_t\mu_s}{\mu_s^2(\mu_s+\mu_t)(\mu_s-\mu_t)}.
\end{align}
Combine (\ref{equ:average_1_remote}), (\ref{equ:exp_y}), (\ref{equ:exp_y_sq}),~(\ref{equ:ex_1_ty_remote}) and (\ref{equ:wy_remote}), (\ref{equ:average_main_remote}) is proven.

\section{Proof of Theorem 3}
Note that the arrival process of the remote computing queue is the same process as the departure process of the tandem of the local server and the transmission channel. The time of computing at the local server or the remote server and the transmission time of channel are iid with exponentially distributed. Thus, partial computing can be viewed as an $GI/M/1$ system.
 Three expectations in equation (\ref{equ:average_1_partial}) need to be calculated for obtaining the average AoI. Since the local computing time $D_i$ and the transmission time $Y_i$ are iid exponentials with average service time $1/\mu_l$ and $1/\mu_t$, respectively. Thus, we can obtain the following equations
\begin{equation}\label{equ:}
E[B_i] = E[D_i+Y_i] =  E[D_i]+E[Y_i]=\frac{1}{\mu_l}+\frac{1}{\mu_t},
\end{equation}
\begin{equation}\label{equ:jointcom_B_exs}
E[B_{i}B_{i-1}]=(E[B])^2=\frac{1}{\mu_l^2}+\frac{1}{\mu_t^2}+\frac{2}{\mu_l\mu_t},
\end{equation}
\begin{equation}\label{equ:jointcom_B_sex}
E[B_{i-1}^2]=E[B^2]=E[(D+Y)^2]=\frac{2}{\mu_l^2}+\frac{2}{\mu_t^2}+\frac{2}{\mu_l\mu_t}.
\end{equation}

Then we calculate $E[T_{i}B_{i-1}]$ in detail. For status update $i$, $T_i$ consists of service time and waiting time similar to (\ref{equ:sys_time}).
The waiting time $W_i$ is in accordance with the system time of the $(i-1)$-th message, $T_{i-1}$, and the inter-arrival time $B_i$. Specially, if $T_{i-1}>B_i$, i.e., message $i$ arrives at the remote computing queue when the $(i-1)$-th message is still in the queue for waiting or is under service, we can get $W_i = T_{i-1}-B_i$. Otherwise, $W_i=0$. Therefore, for message $i$, the waiting time can be written as
\begin{equation}\label{equ:wait_time_joint}
W_i =(T_{i-1}-B_i)^+.
\end{equation}
From (\ref{equ:sys_time}), the term $E[T_{i}B_{i-1}]$ can be written as
\begin{equation}\label{equ:ex_1_joint}
\begin{split}
E[T_{i}B_{i-1}]&\!=E[(W_{i}+S_{i})B_{i-1}]
\!=E[W_{i}B_{i-1}]+E[S_{i}B_{i-1}].
\end{split}
\end{equation}
According to (\ref{equ:sys_time}) and (\ref{equ:wait_time_joint}), we can obtain the term $W_{i}$,
\begin{equation}\label{equ:_joint}
\begin{split}
W_{i}& = (T_{i-1}-B_{i})^+=(W_{i-1}+S_{i-1}-B_{i})^+\\
 &= ((T_{i-2}-B_{i-1})^{+}+S_{i-1}-B_{i})^+.
\end{split}
\end{equation}
Note that the system time $T_{i-2}$ depends on the service time and waiting time of message $(i-2)$, thus it is independent of $S_{i-1}$, $B_{i}$ and $B_{i-1}$. Further more, the system times $T_{i}$ become stochastically identical, i.e., $T=^{st}T_i=^{st}T_{i-1}=^{st}T_{i-2}$ when the system reach a stable state.
The system time $T$'s probability density function for the $GI/M/1$ system is
 \cite{Sztrik2011Basic}
\begin{equation}\label{equ:}
f_{T}(t) = \mu_s(1-\sigma)e^{-\mu_s(1-\sigma)t},~~~ t\geq 0,
\end{equation}
where $\sigma$ satisfies the following equation
 \begin{equation}\label{equ:ls_formu}
\sigma =b^*(\mu_s-\mu_s\sigma)~,
\end{equation}
where $b^*(\cdot)$ is the Laplace-Stieltjes transform of random variable $B$.

For $b>0$, the probability density function of $B_i$ is
 \begin{equation}\label{equ:}
f_B(b):= f_{B_i}(b)=\frac{\mu_l\mu_t}{\mu_t-\mu_l}\left(e^{-\mu_lb}-e^{-\mu_tb}\right),~b>0.
\end{equation}

Then, we have
 \begin{equation}\label{equ:LST}
 \begin{split}
b^*(s)&= E[exp(-sB)]=\int_{0}^{\infty}f_B(b)e^{-sb}db \\
&=\frac{\mu_l\mu_t}{\mu_t-\mu_l}\left[\frac{1}{s+\mu_l}-\frac{1}{s+\mu_t}\right].
\end{split}
\end{equation}
Submitting (\ref{equ:LST}) into (\ref{equ:ls_formu}), we can obtain the following equation
 \begin{equation}\label{equ:ls_joint}
 \begin{split}
\sigma \!\!=b^*(\mu_s\!-\!\!\mu_s\sigma)\!=\!\frac{\mu_l\mu_t}{\mu_t\!\!-\!\mu_l}\left(\!\frac{1}{\mu_s\!\!-\!\mu_s\sigma\!+\!\mu_l}\!-\!\frac{1}{\mu_s\!\!-\!\mu_s\sigma\!+\!\mu_t}\!\right).
\end{split}
\end{equation}
Let $\xi=\mu_s-\mu_s\sigma$, we can get the expression of $\xi$ in equation (\ref{equ:xi}) from equation (\ref{equ:ls_joint}).  Therefore, the system time $T$'s probability density function for the $GI/M/1$ system can be re-expressed as
\begin{equation}\label{equ:}
f_{T}(t) = \xi e^{-\xi t},~~~ t\geq 0.
\end{equation}
We can get the condition expected waiting time $W_{i}$ given $B_{i-1} = b_i$ as
\begin{align}\nonumber
&E[W_{i}|B_{i-1} = b_i]=E[((T_{i-2}-b_{i})^{+}+S_{i-1}-B_{i})^+|B_{i-1} = b_i]\\\nonumber
=&E[((T_{i-2}-b_{i})^{+}+S_{i-1}-B_{i})^+]\\\nonumber
=&\int_{0}^{\infty}f_{T}(t)\int_{0}^{\infty}f_{S}(s)\int_{0}^{\infty}f_{B}(b)\left((t-b_i)^++s-b\right)^+db dsdt,\\\nonumber
=&\int_{0}^{b_i}f_{T}(t)\int_{0}^{\infty}f_{S}(s)\int_{0}^{s}f_{B}(b)(s-b)~dbdsdt\\\nonumber &+\!\int_{b_i}^{\infty}\!\!f_{T}(t)\int_{0}^{\infty}\!\!f_{S}(s)\int_{0}^{t-b_i+s}\!\!f_{B}(b)(t-b_i+s-b)~dbdsdt,\\
=&\frac{\phi-1}{\mu_l}\!-\frac{\varphi+1}{\mu_t}\!+\frac{1}{\mu_s}
+\!\left(\frac{1}{\xi}\!-\frac{\phi}{(\mu_l+\xi)}
+\!\frac{\varphi}{(\mu_t+\xi)}\right)e^{-\xi b_i}, \label{equ:condi_solu_jointcom}
\end{align}
where $\phi=\frac{\mu_t\mu_s}{(\mu_t-\mu_l)(\mu_l+\mu_s)}$ and $\varphi=\frac{\mu_l\mu_s}{(\mu_t-\mu_l)(\mu_t+\mu_s)}$.
Returning to (\ref{equ:ex_1_joint}), be aware that the inter-arrival time $B_{i-1}$ is independent of $S_{i}$. Therefore, (\ref{equ:ex_1_joint}) can be represented as
\begin{equation}\label{equ:ex_1_up}
E[T_{i}B_{i-1}] =E[W_{i}B_{i-1}]+E[S_{i}]E[B_{i-1}],
\end{equation}
where $E[S_{i}]=\frac{1}{\mu_s}$. Then, utilizing the conditional expectation in (\ref{equ:condi_solu_jointcom}), we can have equation (\ref{equ_22_joint})$-$(\ref{equ:joint_1_joint}).
\begin{align}
\label{equ_22_joint}&E[W_{i}B_{i-1}]=\int_{0}^{\infty}b_{i}E[W_{i}|B_{i-1}=b_i]f_{B_i}(b_i)~db_i\\
\label{equ_23_joint}&=\!\int_{0}^{\infty}\!\!E[W_{i}|B_{i-1}=b_i]b_{i}\frac{\mu_l\mu_t}{\mu_t-\mu_l}\left(e^{-\mu_lb_i}-e^{-\mu_tb_i}\right)~db_i\\\nonumber
\label{equ:joint_1_joint}&=\!\frac{\mu_l+\mu_t}{\mu_l\mu_t}\!\left(\frac{1}{\mu_s}\!+\!\frac{\phi-1}{\mu_l}
-\frac{\varphi+1}{\mu_t}\right)+\frac{\mu_l\mu_t}{\mu_t-\mu_l}\left(\frac{1}{(\xi+\mu_l)^2}\right.\\
&\left.-\frac{1}{(\xi+\mu_t)^2}\right)
\left(\frac{1}{\xi}-\!\frac{\phi}{(\mu_l+\xi)}+\frac{\varphi}{(\mu_t+\xi)}\right).
\end{align}
Combining (\ref{equ:average_1_partial}), (\ref{equ:jointcom_B_exs}), (\ref{equ:jointcom_B_sex}),~(\ref{equ:ex_1_up}) and (\ref{equ:joint_1_joint}), we derive the average AoI for partial computing in (\ref{equ:main_joint}). Theorem 3 is proven.

\section{Proof of Theorem 4}
Local computing with deterministic computing time means that the local computing time is constant with $D_i= 1/\mu_l$  for all messages. Then, we have
\begin{equation}\label{equ:exp_d_det}
E[D_{i}D_{i-1}]=D_iD_{i-1} =1/\mu_l^2,
\end{equation}
\begin{equation}\label{equ:exp_d_sq_det}
 E[D_{i-1}^2]=D_{i-1}^2=1/\mu_l^2.
\end{equation}
Submit (\ref{equ:exp_d_det}) and (\ref{equ:exp_d_sq_det}) into (\ref{equ:average_local}), the average AoI for local computing with deterministic computing time can be obtained as
\begin{equation}\label{equ:average_local_age}
\bar{\Delta}= \lim_{\tau\rightarrow \infty}\Delta_\tau =\mu_lE[Q_i]=\frac{3}{2\mu_l}.
\end{equation}
\section{Proof of Theorem 5}
Be aware that the arrival process of the computing queue is equals to the message’s departure process of the transmission channel, which is a Poisson process when utilizing zero-wait policy. And the computing time is deterministic. Thus, the computing queue and the MEC server form an $M/D/1$ system.
In this queuing system, the inter-arrival time $Y_i$ is iid, and follows exponential distribution with mean $E[Y_i] = 1/\mu_t$ and the deterministic service time is $S_i = 1/\mu_s$.
We now calculate $E[T_{i}Y_{i-1}]$ in detail. For status update $i$, $T_i$ has two parts as waiting time and service time, i.e.,
\begin{equation}\label{equ:sys_time_det}
T_{i} = W_i+S_i= W_i+\frac{1}{\mu_s}.
\end{equation}
From (\ref{equ:sys_time_det}), the term $E[T_{i}Y_{i-1}]$ can be written as
\begin{align}\nonumber
E[T_{i}Y_{i-1}]&=E[(W_{i}+S_{i})Y_{i-1}]\\\label{equ:ex_1_remote_det}
&=E[W_{i}Y_{i-1}]+\frac{E[Y_{i-1}]}{\mu_s}.
\end{align}
According to (\ref{equ:sys_time_det}) and (\ref{equ:wait_time}), we can obtain the term $W_{i}$,
\begin{align}\label{equ:}\nonumber
W_{i}& = (T_{i-1}-Y_{i})^+=(W_{i-1}+\frac{1}{\mu_s}-Y_{i})^+\\\nonumber
 &= \left((T_{i-2}-Y_{i-1})^{+}+\frac{1}{\mu_s}-Y_{i}\right)^+\\
 &=\left((W_{i-2}+\frac{1}{\mu_s}-Y_{i-1})^{+}+\frac{1}{\mu_s}-Y_{i}\right)^+.
\end{align}
Since $W_{i-2}$ and $Y_{i-1}$ are independent with each other, the conditional expected waiting time $W_{i}$ given $Y_{i-1} = y_i$ can be obtained as
\begin{align}\nonumber
&E[W_{i}|Y_{i-1} = y_i]\\
=&E[((W_{i-2}+\frac{1}{\mu_s}-Y_{i-1})^{+}+\frac{1}{\mu_s}-Y_{i})^+\big |~Y_{i-1} = y_i]\\\nonumber
=&E[((W_{i-2}+\frac{1}{\mu_s}-y_{i})^{+}+\frac{1}{\mu_s}-Y_{i})^+]\\\nonumber
=&\int_{0}^{\infty}f_{W}(w)\int_{0}^{\infty}f_{Y}(y)\left((w+\frac{1}{\mu_s}-y_i)^{+}+\frac{1}{\mu_s}-y\right)^+dydw,\\\nonumber
=&\int_{0}^{y_i-1/\mu_s}f_{W}(w)\int_{0}^{1/\mu_s}f_{Y}(y)(1/\mu_s-y)~dydw +\\ \label{equ:condi_solu_det} &\!\!\!\int_{y_i-1/\mu_s}^{\infty}\!\!f_{W}(w)\!\!\!\int_{0}^{w-y_i+2/\mu_s}\!\!\!\!f_{Y}(y)(w\!-y_i+2/\mu_s\!-\!y)~\!dydw,
\end{align}
where $f_{W}(w)$ is the probability density function of the waiting time $W$ in an $M/D/1$ queuing system which is obtained as the derivative of the CDF (see \cite{Franx01asimple})
\begin{align}\label{equ:wait_det}
F_W(w) = (1-\rho)\sum_{k=0}^{\lfloor w\mu_s\rfloor}\frac{\rho^k}{k!}(k- w\mu_s)^ke^{\rho(w\mu_s-k)},
\end{align}
where $\rho=\mu_t/\mu_s$.
Utilizing the conditional expectation $E[W_{i}|Y_{i-1} = y_i]$ in (\ref{equ:condi_solu_det}), we can obtain $E[W_iY_{i-1}]$ through the equation (\ref{equ_22}). Having calculated $E[W_iY_{i-1}]$,
we can combine (\ref{equ:average_1_remote}), (\ref{equ:exp_y}), (\ref{equ:exp_y_sq}) and (\ref{equ:ex_1_remote_det}) to obtain the average AoI in remote computing with deterministic computing time.
\section{Proof of Theorem 6}
Note that the arrival process of the remote computing queue is the same process as the departure process of the tandem of the local server and the transmission channel.
The time of computing at the local server and the remote server are deterministic, while the transmission time of channel is exponentially distributed.
Thus, partial computing can be viewed as an $GI/D/1$  system.
 To derive the average AoI, three expectations need to be calculated in equation (\ref{equ:average_1_partial}). As the local computing time $D_i =1/\mu_l$ is constant and the transmission time $Y_i$ is iid, and follows exponential distribution with average service time $1/\mu_t$, we can obtain the following equations
\begin{equation}\label{equ:jointcom_B_sig_det}
E[B_i] = E[D_i+Y_i] =E[\frac{1}{\mu_l}+Y_i]=\frac{1}{\mu_l}+\frac{1}{\mu_t},
\end{equation}
\begin{equation}\label{equ:jointcom_B_exs_det}
E[B_{i}B_{i-1}]=E[(\frac{1}{\mu_l}+Y_i)(\frac{1}{\mu_l}+Y_{i-1})]=\frac{1}{\mu_l^2}+\frac{1}{\mu_t^2}+\frac{2}{\mu_l\mu_t},
\end{equation}
\begin{equation}\label{equ:jointcom_B_sex_det}
E[B_{i-1}^2]=E[(\frac{1}{\mu_l}+Y_i)^2]=\frac{1}{\mu_l^2}+\frac{2}{\mu_t^2}+\frac{2}{\mu_l\mu_t}.
\end{equation}
Then, we need to calculate $E[T_{i}B_{i-1}]$. The term $E[T_{i}B_{i-1}]$ is the same as (\ref{equ:ex_1_joint}). $E[W_{i}B_{i-1}]$ need to be calculated to obtain the average AoI.

The probability density function of $B_i$ is
 \begin{equation}\label{equ:}
f_B(b):= f_{B_i}(b)=\mu_te^{-\mu_t(b-\frac{1}{\mu_l})},~b\geq\frac{1}{\mu_l}.
\end{equation}
The MEC service time is constant with $\frac{1}{\mu_s}$. There is no queueing when $\frac{1}{\mu_l}\geq\frac{1}{\mu_s}$, since the MEC computing time is smaller than the inter-arrival time. Thus, the waiting time is zero. Accordingly, this term $E[W_{i}B_{i-1}]=0$. Therefore, in the case of $\frac{1}{\mu_l}\geq\frac{1}{\mu_s}$, combining (\ref{equ:average_1_partial}), (\ref{equ:ex_1_up}),~and (\ref{equ:jointcom_B_sig_det})$-$(\ref{equ:jointcom_B_sex_det}), we can obtain the average AoI in partial computing  as
\begin{equation}\label{}
\begin{split}
 \bar{\Delta}_p=&\frac{\mu_l\mu_t}{\mu_l+\mu_t}\left(E[S_{i}]E[B_{i-1}]+E[B_{i}B_{i-1}]+\frac{1}{2}E[B_{i-1}^2]\right)\\
 =&\frac{1}{\mu_l+\mu_t}\left(3+\frac{3\mu_t}{2\mu_l}+\frac{2\mu_l}{\mu_t}+\frac{\mu_t}{\mu_s}+\frac{\mu_l}{\mu_s}\right).
\end{split}
\end{equation}
 Then we calculate $E[W_{i}B_{i-1}]$ when $\frac{1}{\mu_l}<\frac{1}{\mu_s}$.
 For status update $i$, $T_i$ comprises the waiting time and service time. According to the equation (\ref{equ:sys_time}) and (\ref{equ:wait_time_joint}),
 we express the waiting time of the $i$-th message as
\begin{align}\nonumber
W_{i}& = (T_{i-1}-B_{i})^+=(W_{i-1}+1/\mu_s-B_{i})^+\\\label{equ:wait_time_joint_det}
 &= ((W_{i-2}+1/\mu_s-B_{i-1})^{+}+1/\mu_s-B_{i})^+.
\end{align}
The conditional expected waiting time $W_{i}$ given $B_{i-1} = b_i$ can be obtained as

\setlength\abovedisplayskip{3pt}
\begin{align}\nonumber
&E[W_{i}|B_{i-1} = b_i]\\
=&E[((W_{i-2}+1/\mu_s-B_{i-1})^{+}+1/\mu_s-B_{i})^+|B_{i-1} = b_i]\\\nonumber
=&E[((W_{i-2}+1/\mu_s-b_i)^{+}+1/\mu_s-B_{i})^+]\\\nonumber
=&\int_{0}^{\infty}\!f_{W}(w)\int_{\frac{1}{\mu_l}}^{\infty}\!f_{B}(b)\left((w+1/\mu_s-b_i)^{+}\!+1/\mu_s-b\right)^+db dw,\\\nonumber
=&\int_{0}^{b_i-1/\mu_s}f_{W}(w)\int_{\frac{1}{\mu_l}}^{\frac{1}{\mu_s}}f_{B}(b)(1/\mu_s-b)~dbdw+\\ &\!\!\int_{b_i-1/\mu_s}^{\infty}\!\!\!f_{W}(w)\!\int_{\frac{1}{\mu_l}}^{w-b_i+\frac{2}{\mu_s}}\!\!\!f_{B}(b)(w-\!b_i+2/\mu_s\!-\!b)~\!\!dbdw, \label{equ:condi_solu_jointcom_det}
\end{align}
\setlength\belowdisplayskip{1pt}
where $f_{W}(w)$ is the probability density function of the waiting time $W$ in the $GI/D/1$ queue system.
As the inter-arrival time is $1/\mu_l$ plus an exponentially distributed random variable while the service time is a constant $1/\mu_s$, the waiting time of this $GI/D/1$ system is the same as that of an $M/D/1$ system with arrival rate $\mu_t$ and deterministic service rate $\mu = \frac{1}{1/\mu_s-1/\mu_l} = \frac{\mu_l\mu_s}{\mu_l-\mu_s}$. Therefore, $f_{W}(w)$ can be obtained as the derivative of the CDF refer to the $M/D/1$ system
\begin{equation}\label{equ:wait}
F_W(w) = (1-\rho)\sum_{k=0}^{\lfloor w \mu\rfloor}\frac{\rho^k}{k!}(k- w\mu)^ke^{\rho(w\mu-k)},
\end{equation}
where $\rho=\mu_t/\mu$.
Returning to (\ref{equ:ex_1_joint}), we note that the inter-arrival time $B_{i-1}$ is independent of $S_{i}$. Therefore, (\ref{equ:ex_1_joint}) can be rewritten as (\ref{equ:ex_1_up}) with $E[S_{i}]=\frac{1}{\mu_s}$.
 Moreover, utilizing the conditional expectation in (\ref{equ:condi_solu_jointcom_det}), $E[W_iB_{i-1}]$ can be calculated numerically according to (\ref{equ_23_partial}). Having calculated $E[W_iB_{i-1}]$, we can calculate the average AoI combining (\ref{equ:average_1_partial}), (\ref{equ:ex_1_up}),~and (\ref{equ:jointcom_B_sig_det})$-$(\ref{equ:jointcom_B_sex_det}).
\begin{align}
E[W_{i}B_{i-1}]=&\int_{\frac{1}{\mu_l}}^{\infty}b_{i}E[W_{i}|B_{i-1}=b_i]f_{B_i}(b_i)~db_i\\
\label{equ_23_partial}
=&\int_{\frac{1}{\mu_l}}^{\infty}E[W_{i}|B_{i-1}=b_i]b_{i}\mu_te^{-\mu_t(b_i-\frac{1}{\mu_l})}~db_i.
\end{align}

\ifCLASSOPTIONcaptionsoff
  \newpage
\fi
\footnotesize
\bibliographystyle{IEEEtran}
\bibliography{references_KQB0111}

\end{document}